\documentclass[a4paper,british]{scrartcl}
\usepackage{lmodern}

\usepackage[T1]{fontenc}
\usepackage[latin9]{inputenc}
\usepackage{babel}
\usepackage{float}
\usepackage{units}
\usepackage{textcomp}
\usepackage{amstext}
\usepackage{graphicx}
\usepackage{setspace}
\usepackage{esint}
\usepackage[authoryear]{natbib}
\usepackage[unicode=true,
 bookmarks=true,bookmarksnumbered=false,bookmarksopen=false,
 breaklinks=false,pdfborder={0 0 1},backref=false,colorlinks=false]
 {hyperref}
\hypersetup{pdftitle={The Impact of Oceanic Heat Transport on the Atmospheric Circulation: a Thermodynamic Perspective},
 pdfauthor={Schroeder Lucarini Lunkeit}}
\usepackage{breakurl}

\makeatletter

\usepackage{babel}

\usepackage{subfigure}

\date{}

\AtBeginDocument{
  
}

\makeatother

\begin{document}

\title{The Impact of Oceanic Heat Transport on the Atmospheric Circulation:
a~Thermodynamic~Perspective}

\begin{onehalfspace}

\author{{\normalsize{Alexander Schröder}}%
\thanks{Email: \protect\href{mailto:alexander.schroeder@studium.uni-hamburg.de}{alexander.schroeder@studium.uni-hamburg.de}\protect \\
 Meteorologisches Institut, Klima Campus, University of Hamburg, Grindelberg
5, 20144 Hamburg%
}{\normalsize{\,, ~Valerio Lucarini}}%
\thanks{Email: \protect\href{mailto:valerio.lucarini@uni-hamburg.de}{valerio.lucarini@uni-hamburg.de}
\protect \\
Tel: +49 (0) 40 42838 9208 \protect \\
Meteorologisches Institut, Klima Campus, University of Hamburg, Grindelberg
7, 20144 Hamburg\protect \\
Department of Mathematics and Statistics, University of Reading, Reading,
RG6 6AX, UK%
}{\normalsize{ ~and Frank Lunkeit}}%
\thanks{Email: \protect\href{mailto:frank.lunkeit@uni-hamburg.de}{frank.lunkeit@uni-hamburg.de}\protect \\
Phone: +49 (0) 40 42838 5073\protect \\
 Meteorologisches Institut, Klima Campus, University of Hamburg, Grindelberg
5, 20144 Hamburg%
}}
\end{onehalfspace}

\maketitle
\begin{onehalfspace}
\thispagestyle{empty}
\end{onehalfspace}

\subsection*{Abstract}
\begin{abstract}
\begin{singlespace}
\noindent The present study investigates how global thermodynamic
properties of the climate system are affected by the changes in the
intensity of the imposed oceanic heat transport in an atmospheric
general circulation model in aqua-planet configuration. Increasing
the poleward oceanic heat transport results in an overall increase
in the surface temperature and a decrease in the equator-to-pole surface
temperature difference as a result of the ice-albedo feedback. Following
the classical ansatz by Stone, the atmospheric heat transport changes
in such a way that the total poleward heat transport remains almost
unchanged. We also find that the efficiency of the climate machine,
the intensity of the Lorenz energy cycle and the material entropy
production of the system decline with increased oceanic heat transport
which suggests that the climate system becomes less efficient and
turns into a state of reduced entropy production, as the enhanced
oceanic transport performs a stronger large-scale mixing between geophysical
fluids with different temperature, thus reducing the availability
in the climate system and bringing it closer to a state of thermal
equilibrium. 
\end{singlespace}

\newpage{}

\setcounter{page}{2}\pagenumbering{arabic}
\end{abstract}

\section{Introduction}

The climate is a forced and dissipative non-equilibrium system, which,
neglecting secular trends, can be considered in steady state, i.e.
its statistical properties do not depend on time, and on the global
scale, the energy and entropy budgets are closed. The positive budget
of incoming over outgoing radiation at the top of the atmosphere in
the tropical regions is compensated by a negative budget in the high
latitudes. The large-scale geophysical fluids transport the excess
of energy from low to high latitudes. The entropy budget is achieved
in such a way that the sum of the integrated incoming entropy flux
due to the solar high frequency photons plus the entropy generated
by irreversible processes in the geophysical fluids and as a result
of the the absorption and emission of photons are compensated by the
radiation to space of low frequency photons. Most of the entropy production
results from optical processes, while a smaller portion - referred
to as material entropy production - is related to the irreversible
processes in terms of the geophysical fluids \citep{Kleidon2005}.
So the Earth is, in contrast to a system that is isolated and, therefore,
maintaining a state of equilibrium, a thermodynamic system that exchanges
energy and entropy with space \citep{Ambaum2010}.

The large-scale fluid climatic motions result from the conversion
of available potential energy - due to the inhomogeneous absorption
of solar radiation, with positive correlation between heating and
temperature patterns - into kinetic energy, through instabilities
coming, typically, from the presence of temperature gradients \citep{Lorenz1954}.
Such instabilities tend to reduce the same temperature gradients they
feed upon, by mixing fluid masses. The kinetic energy is then dissipated
inside the system. In steady state conditions, the production of available
potential energy, its conversion to kinetic energy, and the dissipation
of kinetic energy have the same average rate, which corresponds to
the intensity of the Lorenz \citeyearpar{Lorenz1954,Lorenz1967} energy
cycle. The closure of such a thermodynamical/dynamical problem amounts
to a self-consistent theory of climate.

\begin{onehalfspace}
Fuelled by the presence of temperature gradients, the climate system
can be interpreted as a thermal engine that converts potential into
mechanical energy \citep{Peixoto}. Recently, using tools of macroscopic
nonequilibrium thermodynamics, a line has been drawn connecting a
measure of the efficiency of the climate system, the spatio-temporal
variability of its heating and temperature fields, the intensity of
the Lorenz energy cycle, and the material entropy production \citep{Johnson2000,Lucarini2009,Lucarini}.

The role of oceanic heat transport in the climate system is a central
aspect of climate dynamics and has drawn a great attention in the
field of climate science. \citet{Herweijer2005} observed that the
presence of oceanic heat transport results into an overall warming
of the surface. \citet{Barreiro2011a} have shown that an increase
in oceanic heat transport raises the global mean temperature by decreasing
the albedo due to reduced sea-ice extent and marine stratus cloud
cover and by increasing the greenhouse effect through a moistening
of the atmosphere. \citet{Rose2013} studied the role of oceanic heat
transport with an idealised aqua-planet GCM. They focused on the problem
of clarifying how oceanic heat transport affects the meridional surface
temperature field. 
\end{onehalfspace}

\citet{Stone1977} argued that the total poleward heat transport,
defined as the sum of atmospheric and oceanic transport, is to a good
approximation set by the planetary albedo and astronomical parameters.
\citeauthor{Stone1977} states that the transport regime is weakly
sensitive to the presence of oceans, mountains or the hydrological
cycle, i.e. detailed atmospheric and oceanic processes that impact
the transport mechanism. Experiments with coupled GCMs have shown
an almost complete compensation by atmospheric heat transport for
variations in oceanic heat transport \citep{Manabe1969,Covey1988,Enderton2009}.

\begin{onehalfspace}
In the present study we explore the impact of the oceanic heat transport
on macroscale thermodynamic quantities of the climate system. The
aim is to characterise global nonequilibrium properties of the climatic
machine in terms of their thermodynamical steady-state response to
the change in the oceanic heat transport. Therefore, this work complements
previous findings focusing on the investigation of the impacts of
mean climate properties and circulation patters. We first look into
the response of the meridional atmospheric heat transport, in order
to test the hypothesis of ocean/atmosphere compensation, proposed
in various forms by e.g. \citet{Bjerknes1964} and \citet{Stone1977}.
We then move from considering energy fluxes to looking at energy transformation,
i.e. we investigate how changes in the ocean transport impact the
intensity of the Lorenz \citeyearpar{Lorenz1954} energy cycle, and
link this to changes in the spatio-temporal variability of the temperature
in the atmosphere, by studying the properties of the effective warm
and cold reservoirs constructed according to the theory proposed in
\citet{Johnson2000}, \citet{Lucarini2009} and \citet{Lucarini2010},
so allowing the definition of a measure of the efficiency of the climate
system. Finally, we will direct our attention to measuring the irreversibility
of the climate system and study the impact of changing oceanic transport
intensity on the material entropy production of the climate system.
Our analysis tries to frame specific climatic processes of general
relevance into a general physical framework, trying to advance the
understanding of the climate as a non-equilibrium, forced and dissipative
macroscopic system. 
\end{onehalfspace}

As a first step into this way of looking at the coupling between atmosphere
and ocean, we consider a simplified yet physically relevant modelling
set-up. We use a climate model of intermediate complexity, PlaSim
\citep{Fraedrich2005a}, which features a simplified yet reasonable
representation of the 3D dynamics of the atmosphere and of its interactions
with land and ocean boundary and surface layers. Instead, the representation
of the ocean processes is severely simplified, as no explicit description
of dynamic processes is given. The ocean rather provides prescribed
lower boundary conditions (sensible and latent heat fluxes, albedo)
for the atmosphere above. While this is an obvious limitation in terms
of realism, such a setting allows for flexibly modulating the atmosphere/ocean
interaction.

\begin{onehalfspace}
The paper is organized as follows. In section 2 we summarise the theoretical
background on non-equilibrium thermodynamical properties of the climate
system. In section 3 we describe the properties of the numerical model
considered in this study and present the set of experiments we have
performed. In section 4 we present the main results of our work. The
summary of our findings and the related discussion are presented in
section 5. 
\end{onehalfspace}

\section{Theoretical framework}

\begin{onehalfspace}
We wish to highlight two important aspects of the geophysical fluids
of the climate system. First, they transport heat from regions featuring
net positive energy budget at the top of the atmosphere (low latitudes)
to regions where such budget is negative (high latitudes), thus reducing
the temperature gradient between equator and poles \citep{Peixoto,Lucarini2011a}.
Secondly, they perform on the average a net positive work due to the
positive correlation between temperature and heating fields. Such
work is used to uphold the kinetic energy of the global circulation
against the frictional dissipation \citep{Peixoto}. The general circulation
of the atmospheric system arises from the conversion of available
potential energy into kinetic energy (e.g. atmospheric motions), as
introduced in the formulation of the energy cycle in \citet{Lorenz1954,Lorenz1967}.
If the climate system is at statistical steady state, the rate of
generation of available potential energy $\dot{G}$, the rate of conversion
of potential into kinetic energy $\dot{W}$, and the dissipation rate
of kinetic energy $\dot{D}$ are equal when averaged over a long period
of time (e.g. several years), so that $\overline{\dot{G}}=\overline{\dot{W}}=\overline{\dot{D}}>0$,
where the bar indicates the operation of time averaging. This allows
for characterising the strength of the Lorenz energy cycle in several
ways. Let us briefly recapitulate, following \citet{Johnson2000}
and \citet{Lucarini2009}, some thermodynamic ideas we will use throughout
the paper.

Let $\Omega$ be the volume domain of the climate system and $\dot{Q}$
be the local heating rate due to frictional dissipation and convergence
of heat fluxes including radiative, sensible and latent heat components.
At each instant $t$ we divide $\Omega$ into two subsections, so
that $Q(x,t)>0,\, x\in\Omega^{+}$ defining $Q^{+}$, and $Q(x,t)<0,\, x\in\Omega^{-}$
for $Q^{-}$ respectively. We wish to remark that the domains $\Omega^{+}$
and $\Omega^{-}$ are time dependent. Integrating the two heating
components results in: $\int_{\Omega^{+}}\rho\dot{Q}^{+}\mathrm{d}V+\int_{\Omega^{-}}\rho\dot{Q}^{-}\mathrm{d}V=\dot{\Phi}^{+}+\dot{\Phi}^{-}.$
\citet{Johnson2000} and \citet{Lucarini2009} show that the time
average $\overline{\dot{\Phi}^{+}}+\overline{\dot{\Phi}^{-}}$ gives
the rate of generation of available potential energy, so that: 
\begin{equation}
\overline{\dot{W}}=\overline{\dot{\Phi}^{+}}+\overline{\dot{\Phi}^{-}}.
\end{equation}
The efficiency of the climate machine can now be expressed as: 
\begin{equation}
\eta=\frac{\overline{\dot{\Phi}^{+}}+\overline{\dot{\Phi}^{-}}}{\overline{\dot{\Phi}^{+}}}.\label{eq:efficiency}
\end{equation}
This expression represents the ratio for the work output $\overline{\dot{\Phi}^{+}}+\overline{\dot{\Phi}^{-}}$
to the heat input $\overline{\dot{\Phi}^{+}}$. At each instant one
defines the quantities $\Sigma^{+(-)}=\nicefrac{\int_{\Omega}^{+(-)}\rho\dot{Q}^{+(-)}}{T}$,
which are the instantaneous entropy sources and sinks in the system.
As explained in \citet{Johnson2000} and \citet{Lucarini2009}, we
have that $\overline{\dot{\Sigma}^{+}}+\overline{\dot{\Sigma}^{-}}=0$.
We can then introduce the scale temperatures $\Theta^{+}=\nicefrac{\overline{\dot{\Phi}^{+}}}{\overline{\dot{\Sigma}^{+}}}$
and $\Theta^{-}=\nicefrac{\overline{\dot{\Phi}^{-}}}{\overline{\dot{\Sigma}^{-}}}$,
so that equation \ref{eq:efficiency} can be rewritten as $\eta=\frac{\Theta^{+}-\Theta^{-}}{\Theta^{+}}$,
where $\Theta^{+}>\Theta^{-}$. 
\end{onehalfspace}

Hence, the motion of the general circulation of the system can be
sustained against friction because zones being already relatively
warm absorb heat whereas the relatively low temperature zones are
cooled.

The Lorenz energy cycle can thus be seen as resulting from the work
of an equivalent Carnot engine operating between the two (dynamically
determined) reservoirs at temperature $\Theta^{+}$ and $\Theta^{-}$.
Yet, the climate is far from being a perfect engine, as many irreversible
processes take place; nonetheless, a Carnot-equivalent picture can
be drawn as described.

\begin{onehalfspace}
Let us now delve into such irreversible processes. In the climate
system two rather different sets of processes contribute to the total
entropy production \citep{Peixoto,Goody1999,Ambaum2010}. The first
set of processes is responsible for the irreversible thermalisation
of photons emitted near the Sun's corona at roughly 5800\,K, absorbed
and then re-emitted at much lower temperatures, typical of the Earth's
climate ($\sim$255\,K). This gives the largest contribution to the
total average rate of entropy production for the Earth system of about
900\,mW\,m$^{-2}$\,K$^{-1}$ \citep{Peixoto,Ambaum2010}. The
remaining contribution is due to the processes responsible for mixing
and diffusion inside the fluid component of the Earth system, and
for the dissipation of kinetic energy due to viscous processes. This
constitutes the so-called material entropy production, and is considered
to be the entropy related quantity of main interest as far as the
properties of the climate system are concerned. Further relevant research
on entropy production in the climate system treating also the geochemical
and radiative contribution to entropy production can be found in \citet{Kleidon2009}
and \citet{Wu2010} respectively.

The entropy budget of geophysical fluids at steady state, following
\citet{Goody1999,Lucarini}, is given by: 
\begin{equation}
\overline{\dot{S}(\Omega)}=\int_{\Omega}\rho\left(\overline{\frac{\dot{q}_{rad}}{T}}+\overline{\dot{s}_{mat}}\right)\mathrm{d}V=0,\label{eq:ent-budg}
\end{equation}

where $\dot{q}_{rad}$ indicates the heating rate by the convergence
of radiative fluxes, $T$ is the local temperature at which the energy
is gained or lost, while $\dot{s}_{mat}$ represents the density of
entropy production associated with the irreversibility of processes
involving the fluid medium. Equation \ref{eq:ent-budg} represents
the entropy budget and states that in a steady state the radiative
entropy source must be balanced by the rate of material entropy production
$\dot{S}_{mat}$ due to material irreversible processes. See a detailed
discussion of this aspect in \citet{Lucarini2014}, where the contributions
to the material entropy production at various spatial and temporal
scales are discussed. 
\end{onehalfspace}

In a steady-state climate the material entropy production $\dot{S}_{mat}(\Omega)$
can be expressed in general terms as:

\begin{equation}
\overline{\dot{S}_{\mathrm{mat}}(\Omega)}=\int_{\Omega}\rho\overline{\dot{s}_{mat}}\mathrm{d}V=\int_{\Omega}\overline{\frac{\varepsilon^{2}}{T}}\mathrm{d}V+\int_{\Omega}\overline{(\vec{F}_{sens}+\vec{F}_{lat})\cdot\vec{\nabla}\frac{1}{T}}\mathrm{d}V=-\int_{\Omega}\rho\overline{\frac{\dot{q}_{rad}}{T}}\mathrm{d}V,\label{eq:mat-entropy}
\end{equation}

\begin{onehalfspace}
where $\overline{\dot{s}_{mat}}$ is the time averaged density of
entropy production due to the following irreversible processes inside
the medium: dissipation of kinetic energy ($\varepsilon^{2}$ is the
specific dissipation rate) and turbulent transport of heat down the
temperature gradient ($\vec{F}_{sens}$ and $\vec{F}_{lat}$, being
the sensible and latent turbulent heat fluxes, respectively). 
\end{onehalfspace}

One needs to underline that a more refined treatment of the entropy
production related to the hydrological cycle has been proposed by
e.g. \citet{Pauluis2002}, \citet{Pauluis2002a} and \citet{Romps2008}.
Nonetheless, as discussed in \citet{Lucarini2014a}, the overall contribution
of the entropy production due to the hydrological cycle can be reconstructed
to a high degree of accuracy also in the simplified method proposed
here.

Note that one can compute the entropy production as:

\begin{equation}
\overline{\dot{S}_{\mathrm{mat}}(\Omega)}=\int_{\Omega}\overline{\frac{\varepsilon^{2}}{T}}\mathrm{d}V+\int_{\Omega}\overline{\frac{\vec{-\nabla\cdot}(\vec{F}_{sens}+\vec{F}_{lat})}{T}}\mathrm{d}V+\int_{\partial\Omega}\overline{\frac{\vec{F}_{sens}+\vec{F}_{lat}}{T}}\cdot\hat{n}\mathrm{d}S,\label{eq:mat-entropy-1}
\end{equation}

where the first term is unchanged, the second terms describes the
entropy gain and loss due to heating and cooling by convergence of
sensible and latent heat fluxes, and the last term is the net entropy
flux across the boundaries of $\Omega$. If one consider the atmospheric
domain as $\Omega$, such term becomes equal to the integral at surface
of the ratio between the sum of the sensible and of the latent heat
flux divided by the surface temperature. Equation \ref{eq:mat-entropy-1}
represents the way entropy production is typically computed in numerical
models. If one considers the whole climate system as $\Omega$, the
boundary terms disappear. Nonetheless, another term proportional to
a Dirac's delta at $z=z_{surf}=0$ appears, resulting from the divergence
of the turbulent flux due to the net evaporation at surface. If we
integrate over $\Omega$, the contribution of this term is exactly
the same as in the case where $\Omega$ corresponds to the atmosphere
only. In other terms, our simplified representation of the ocean is
such that all the entropy is produced in the atmosphere.

We can now separate in equation \ref{eq:mat-entropy} - or, equivalently,
in equation \ref{eq:mat-entropy-1} the first term from the rest,
so that, following \citet{Lucarini2009}, the material entropy production
can be expressed as:

\begin{onehalfspace}
\begin{equation}
\overline{\dot{S}_{mat}(\Omega)}=\overline{\dot{S}_{min}(\Omega)}+\overline{\dot{S}_{exc}(\Omega)},\label{eq:exc-min}
\end{equation}

where $\overline{\dot{S}_{min}(\Omega)}$ is the minimum value of
entropy production compatible with the presence of average dissipation
rate $\int_{\Omega}\epsilon^{2}\mathrm{d}V$, while $\overline{\dot{S}_{exc}(\Omega)}$
is the excess of entropy production with respect to such minimum.
One can associate $\overline{\dot{S}_{min}}$ exactly with the term
in equation \ref{eq:mat-entropy} related to the dissipation of kinetic
energy, while $\overline{\dot{S}_{exc}}$ can be identified with the
sum of the other two terms. 
\end{onehalfspace}

If we take the ratio of the two terms on the right-hand side in equation
\ref{eq:exc-min}, we have that\@. 
\begin{equation}
\alpha=\frac{\overline{\dot{S}_{exc}(\Omega)}}{\overline{\dot{S}_{min}(\Omega)}}\approx\frac{\int_{\Omega}\overline{\vec{(F}_{sens}+\vec{F}_{lat})\cdot\vec{\nabla}\frac{1}{T}}\mathrm{d}V}{\int_{\Omega}\overline{\frac{\varepsilon^{2}}{T}}\mathrm{d}V},\label{eq:alpha}
\end{equation}

where $\alpha$ is the degree of irreversibility \citep{Lucarini2009}
and determines the ratio between the contributions to entropy production
by down-gradient turbulent transport and by viscous dissipation of
mechanical energy. If this ratio is close to zero ($\alpha\rightarrow0$),
all the production of entropy is exclusively caused by unavoidable
viscous dissipation. Pursuing this further, if the turbulent heat
transport in the system from high to low temperature regions is enhanced,
then entropy production is also increased. However, if the turbulent
heat transport down the temperature gradient is maximised, the efficiency
declines due to the reduction of the temperature difference between
the warm and cold reservoir.

\section{Experimental design}

\begin{onehalfspace}
The Planet Simulator (\href{http://www.mi.uni-hamburg.de/fileadmin/files/forschung/theomet/planet_simulator/downloads/PS_ReferenceManual.pdf}{PlaSim};
\citealp{Fraedrich2005a}), a climate model of intermediate complexity
and freely available at~\href{http://www.mi.uni-hamburg.de/index.php?id=216}{www.mi.uni-hamburg.de/plasim},
is applied and used within an idealised aqua-planet configuration.
PlaSim consists of a dynamical core that solves primitive equations
numerically. Unresolved processes are parametrised for: long- and
short-wave radiation \citep{Sasamori1968,Lacis1974}, moist \citep{Kuo1965}
and dry convection, cloud formation \citep{Stephens1978,Stephens1984,Slingo1991}
and large-scale precipitation, latent and sensible heat boundary layer
fluxes, horizontal and vertical diffusion \citep{Louis1979,Laursen1989,Roeckner1992}.

The main goal of this work is to study the sensitivity of the nonequilibrium
thermodynamical properties of the climate to the intensity of the
oceanic heat transfer. Inspired by \citet{Rose2013}, we have performed
climate simulations in which the oceanic heat transport can be parametrically
modulated. Peaks of the oceanic heat flux range from $0.0$\,PW over
present-day value, which is roughly $1.0$\,PW - $2.0$\,PW \citep{Trenberth2001},
up to $4.0$\,PW.

The setup includes a global slab ocean with a $60$\,m mixed-layer
including a thermodynamic sea-ice model. Oceanic heat transport is
controlled by a simple analytical expression to be explained below.
The astrophysical parameters are adjusted to present-day values for
planet Earth except that eccentricity is set to zero.

The series of experiments is performed in a setup of 5 vertical layers
and T\,31 spectral resolution, corresponding to a Gaussian grid resolution
of about $3.75{^{\circ}}\times3.75{^{\circ}}$. This particular setup
is an extreme simplification. However, basic physical elements crucial
for simulating climate features are retained following \citealp{Rose2013}.
The analytical function used to control the prescribed oceanic heat
transport $\Psi$ as a function of geographical latitude $\phi$ is

\begin{equation}
\Psi(\phi)=\Psi_{\mathrm{amp}}\mathrm{sin}(\phi)\mathrm{cos^{2\mathit{N}}}(\phi),\label{eq:oht-rose}
\end{equation}

where $\Psi_{\mathrm{amp}}$ is the amplitude in units of PW and $N$
is the scale parameter, which is held constant by the value $N=2$
for all simulations. $\Psi_{\mathrm{max}}$ is the peak oceanic heat
transport and $\phi_{\mathrm{max}}$ is the latitude at which the
oceanic heat transport $\Psi$ reaches a maximum. For $N=2$ the poleward
transport is maximised at about $27\text{\textdegree}$\,N/S.

Each simulation is performed using a different value of $\Psi_{\mathrm{max}}$
which varies between $0.0$\,PW (control run) to $4.0$\,PW with
an increment of $0.5$\,PW, i.e. $\mathrm{\Psi}_{\mathrm{max}}=\left\{ 0.0,0.5,...,4.0\right\} $\,PW.

\begin{figure}[H]
\begin{centering}
\subfigure{\label{fig:OHT}\includegraphics[height=160bp]{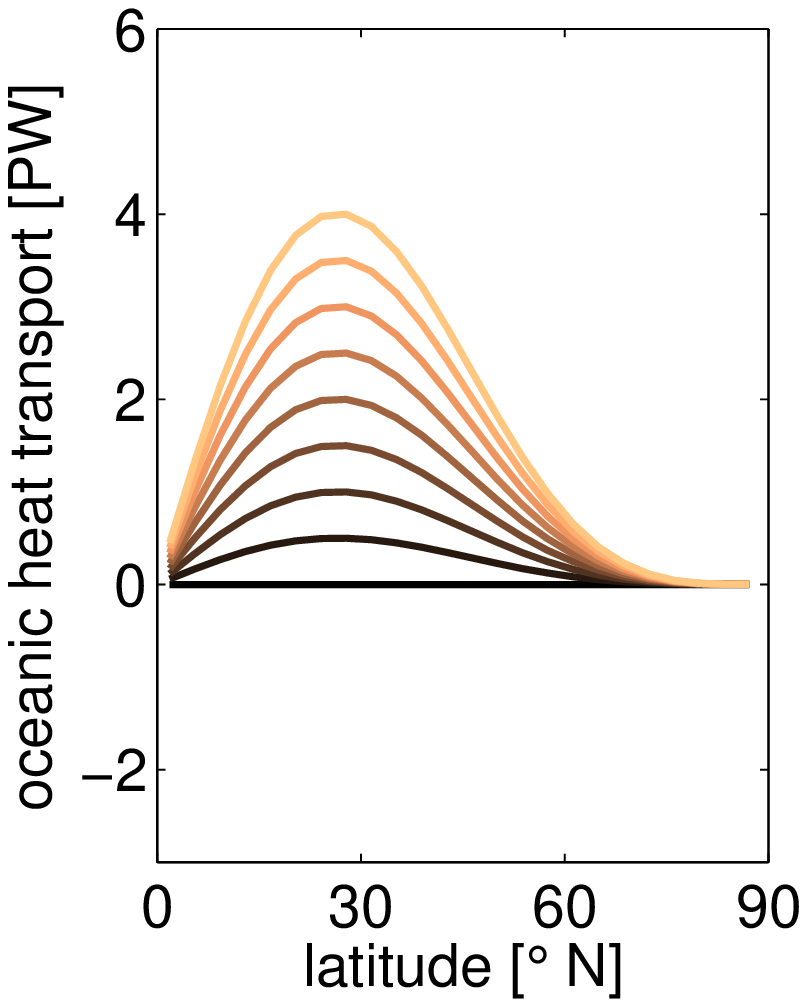}}~~~~~~\subfigure{\label{fig:q-flux}\includegraphics[height=160bp]{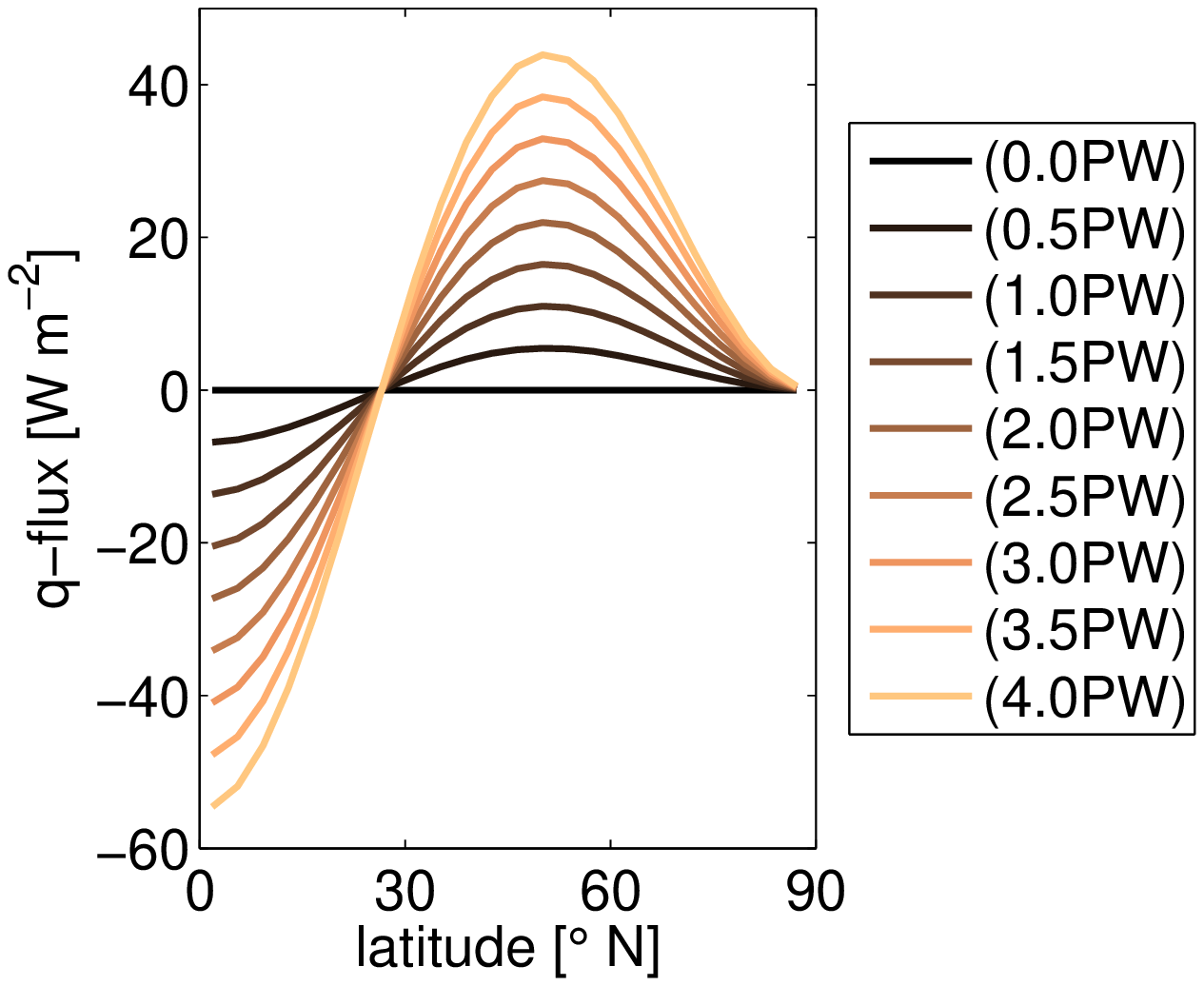}} 
\par\end{centering}

\caption{Oceanic heat transport (a) and q-flux (b) as a function of latitude
for all simulations.\label{fig:Oceanic-heat-transport}}
\end{figure}

The oceanic heat transport and its convergence, the so-called q-flux
\citep{Rose2013}, which determines the oceanic heat transport, are
displayed in figure \ref{fig:OHT}. The q-flux implemented in the
model is given the characteristics of being zonally symmetric and
steady in time for all simulations. As shown in figure \ref{fig:q-flux},
the q-flux is negative in the tropics (heat uptake into the ocean)
and positive in the mid and high latitudes (heat release into the
atmosphere). As a consequence of the heat uptake and release at the
ocean's surface, there is a maximum transport of heat just in between
those regions of heat absorption and release, as displayed in figure
\ref{fig:OHT}.

Each model run consists of 100\,years to ensure the final 30\,years,
that are used for diagnostics, are free from any influence of the
model's transient phase at the beginning of each run. 
\end{onehalfspace}

\section{Results}

\begin{figure}[!h]
\begin{centering}
\subfigure{\label{fig:AHT}\includegraphics[height=160bp]{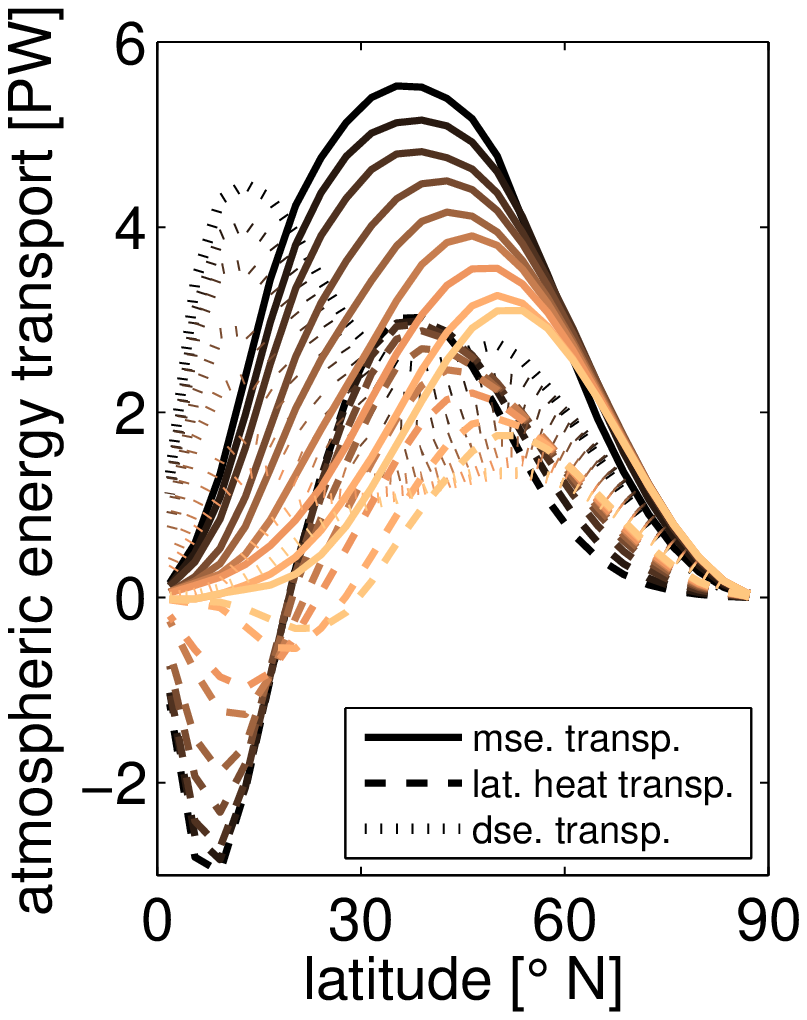}}~~~~~~\subfigure{\label{fig:THT}\includegraphics[height=160bp]{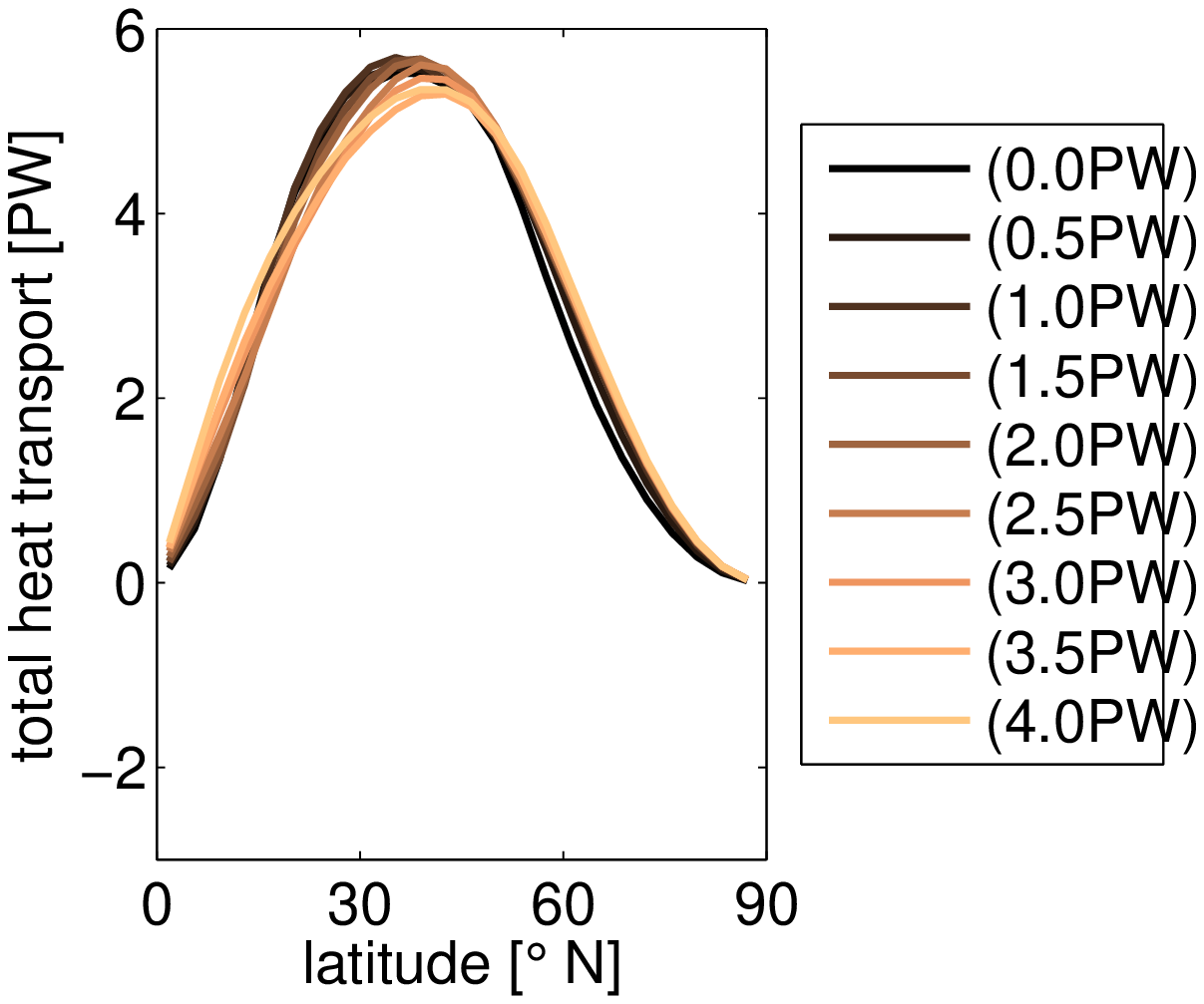}} 
\par\end{centering}

\caption{Time-averaged meridional moist static energy transport (mse. transp.)
with latent heat transport (lat. heat transp.) and dry static energy
transport (dse. transp.) (a), and the total energy transport (b) as
oceanic heat transport is increased.\label{fig:Time-averaged-meridional}}
\end{figure}

\begin{onehalfspace}
We study the atmospheric response in the meridional heat transport
for different scenarios of the q-flux in figure \ref{fig:AHT} by
looking into the time-averaged and vertically integrated zonal mean
transport of moist static energy.

As well known, the geophysical fluids transport energy from the low
to the high latitude regions, the effect being that the meridional
temperature gradient decreases and entropy is produced \citep{Peixoto,Ozawa2003}.
We find that the peak amplitude of the atmospheric energy transport
decreases linearly with the value of the peak oceanic energy transport
$\Psi_{\mathrm{max}}$. For every $0.5$\,PW increase in oceanic
energy transport peak $\Psi_{\mathrm{max}}$, the maximum intensity
of the atmospheric energy transport decreases by about $0.3$\,PW.
When increasing the peak oceanic heat transport from $\Psi_{\mathrm{max}}=0.0$\,PW
to $\Psi_{\mathrm{max}}=4.0$\,PW, we observe a latitudinal relocation
of the peak atmospheric heat flux from $35\text{\textdegree}$ to
$53\text{\textdegree}$, with the effect of shifting the maximum of
the baroclinic activity. As a consequence of the latitudinal shift
as well as the decrease in the maximum intensity, the atmospheric
heat transport features a decrease in the tropics and subtropics because
here the oceanic transport is increased and thus dominates the total
transfer out of the tropics. The strongest changes in the latitudinal
profile of the atmospheric transport take place in the low and mid
latitudes. Between $16$\textdegree{} and $28$\textdegree{}\,N/S
deviations in atmospheric heat transport from the control run are
largest.

The total meridional energy transport, which is computed as the sum
of oceanic and atmospheric energy transport, is almost unchanged for
increased $\Psi_{\mathrm{max}}$ (see figure \ref{fig:THT}). This
result confirms the statement in \citet{Stone1977} about the insensitivity
of the magnitude of the total transport to internal parameters (e.g.
meridional temperature gradient). Outgoing longwave radiation balances
the changes in absorbed short-wave radiation, so that as a result,
the net radiative forcing at top of the atmosphere is roughly unchanged.
Similar results have been presented in \citet{Rose2013}.

\begin{figure}[!t]
\begin{centering}
\includegraphics[height=170bp]{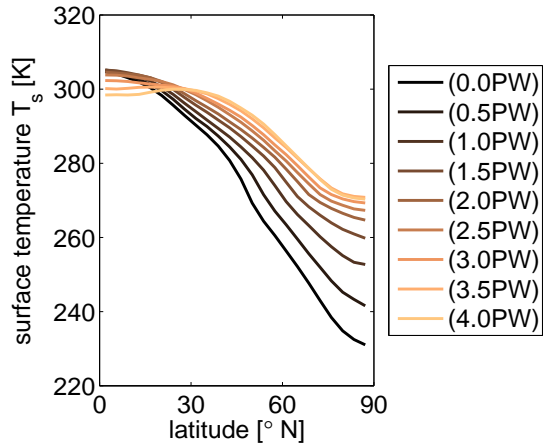} 
\par\end{centering}

\caption{Meridional time-averaged surface temperature profile as a function
of oceanic heat transport.\label{fig:Meridional-surface-temperature}}
\end{figure}

Surface temperature is indeed a quantity of fundamental interest for
climate studies. Figure \ref{fig:Meridional-surface-temperature}
shows the time-averaged zonal mean surface temperature profile with
increased oceanic heat transport. The equator-to-pole temperature
gradient $\Delta T=T_{s,eq}-T_{s,pole}$ is larger than $70$\,K
in the control run. This gradient largely decreases as oceanic heat
transport increases. On average $\Delta T$ declines by $5.5$\,K
for every $0.5$\,PW increase in $\Psi_{\text{max}}$. The gradient
reduction is mainly due to a temperature increase at the poles, except
for oceanic heat transport larger than $2.5$\,PW at which the temperature
decrease at the equator becomes more relevant.

We note a temperature maximum in the subtropics on both hemispheres
for the runs with $\Psi_{\text{max}}=3.5$\,PW and $\Psi_{\text{max}}=4.0$\,PW,
where a small reversed temperature gradient between the deep tropics
and subtropics develops.

\begin{figure}[H]
\begin{centering}
\includegraphics[height=180bp]{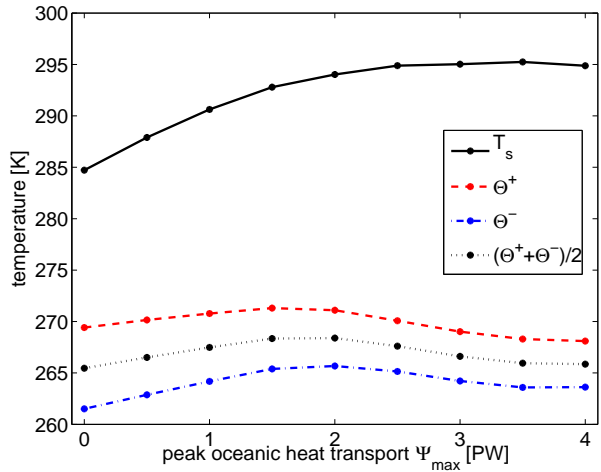} 
\par\end{centering}

\caption{Time average of the global mean surface temperature $T_{s}$ and of
the temperature of the warm ($\Theta^{+}$) and the cold ($\Theta^{-}$)
pool.\label{fig:warm_cold_temperature}}
\end{figure}

The time average of the global mean temperature at the surface $T_{s}$
features a positive sensitivity on the increased $\Psi_{\text{max}}$
(see figure \ref{fig:warm_cold_temperature}). The climate system
features a global warming at the surface of roughly $10$\,K for
the whole range of $\Psi_{\text{max}}$. This is because increased
near-surface heat transport in the northern regions reduces the sea-ice
extent which feeds into the positive ice-albedo feedback, with the
ensuing increase of the global surface temperature. \\
 Let's now shift our attention to the two quantities $\Theta^{+}$
and $\Theta^{-}$, which characterise the warm and cold reservoirs
of the climate engine. Qualitatively, the two temperatures behave
similarly when $\Psi_{\text{max}}$ is changed. We can classify three
temperature regimes: i) $\Psi_{\text{max}}<2.0$\,PW atmospheric
warming, ii) $2.0\,\mathrm{PW}\leq\Psi_{\text{max}}\leq3.5$\,PW
atmospheric cooling, and iii) $\Psi_{\text{max}}>3.5$\,PW weak sensitivity.
We observe a higher sensitivity of $\Theta^{-}$ than $\Theta^{+}$
for i) which is generally due to the amplified polar warming. The
difference between $\Theta^{+}$ and $\Theta^{-}$, denoted as $\Delta\Theta$,
decreases with increasing $\Psi_{\text{max}}$. We now try to find
a rationale of why increases in the imposed oceanic heat transport
cause a reduction of the temperature difference between the warm and
cold reservoir, thereby implying a decrease in the atmospheric efficiency
of the climate engine. Interestingly, the difference between $T_{s}$
and the average of $\Theta^{-}$ and $\Theta^{+}$ increases with
$\Psi_{\text{max}}$, especially for $\Psi_{\text{max}}\leq3.0$\,PW,
indicating a reduction in the stability of the atmosphere. This is
understood by considering that larger oceanic transports lead to stronger
warming at low levels in the mid and high latitudes, which, as suggested
by figure \ref{fig:Time-averaged-meridional}, must be compensated
by a weaker heat transport aloft.

\begin{figure}[!th]
\begin{centering}
\includegraphics[height=280bp]{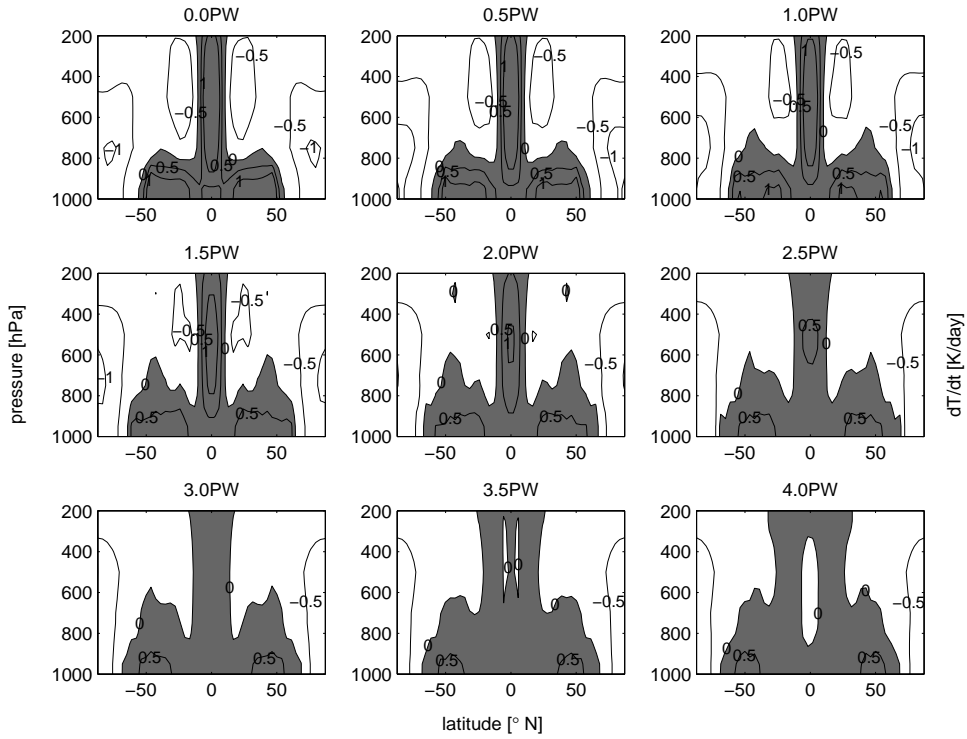} 
\par\end{centering}

\caption{Zonally averaged mean heating rates in the atmosphere for oceanic
heat transport ranging from $0.0$\,PW (upper left panel) to $4.0$\,PW
(low right panel), where grey-shaded areas indicate positive and white
areas negative heating rates in {[}$\nicefrac{\mathrm{K}}{\mathrm{day}}${]}.\label{fig:x-z_heating-rates}}
\end{figure}

The diabatic heating processes constitute the sources and sinks of
internal energy for the atmosphere and play a decisive role in the
generation and destruction of available potential energy \citep{Peixoto}.
Those processes are displayed as the time- and zonal-averaged diabatic
heating rates $\nicefrac{dT_{a}}{dt}$ (see figure \ref{fig:x-z_heating-rates}).
The heating rate is calculated as the sum over all diabatic heating
effects including heating or cooling by the response of radiative
heat fluxes, sensible and latent heat fluxes and vertical diffusion.
While $\Theta^{+}$ and $\Theta^{-}$ are defined using the time and
space dependent heating fields, as described above, inspecting the
time and zonal averages of the heating patterns is useful for understanding
how available potential energy is generated.

Simulations with $0.5\mathrm{\, PW}\leq\Psi_{\text{max}}\leq1.5$\,PW
show diabatic warming in the deep tropics in the mid troposphere and
in the subtropical low troposphere, whereas diabatic cooling occurs
in the mid and high troposphere of the subtropics and in polar as
well as subpolar regions. Positive heating in the tropical and subtropical
regions is dominated by the contribution of latent heat fluxes, in
particular, heating through convective precipitation (not shown here).
In the mid to high latitude regions large-scale precipitation contributes
towards a positive heating. Diabatic cooling, on the other hand, is
mostly caused by outgoing longwave radiation and to a moderate extent
by the conversion process from rain to snow mostly in the subtropical
regions.

We see an extension of the area of positive heating in the mid latitudes
towards the poles in the lower troposphere as well as in the equatorial
mid and upper troposphere for larger values of $\Psi_{\text{max}}$.
The poleward migration of the positive heating pattern in the mid-latitudes
is closely related to the poleward shift of the atmospheric latent
heat transport. The area of positive heating broadens in height at
latitudes around $50\text{\textdegree}$. Since the positive heating
patterns (relevant for defining $\Theta^{+}$) in the mid latitudes
extend in height and is, in addition, stretched poleward, lower temperatures
are considered in the quantity of $\Theta^{+}$, which explains the
smaller sensitivity of $\Theta^{+}$ than of $\Theta^{-}$ for $0.0\mathrm{\, PW}\leq\Psi_{\text{max}}\leq1.5$\,PW
in figure \ref{fig:warm_cold_temperature}. By implication, the warming
effect at polar latitudes causes the sensitivity of $\Theta^{-}$
to be larger than of $\Theta^{+}$. For $\Psi_{\text{max}}\geq2.0$\,PW
the sensitivity of both, $\Theta^{+}$ and $\Theta^{-}$, is negative
since large parts of the tropical high and mid troposphere cools.

\end{onehalfspace}

\begin{figure}[H]
\begin{centering}
\subfigure{\label{fig:reg1}\includegraphics[clip,height=150bp]{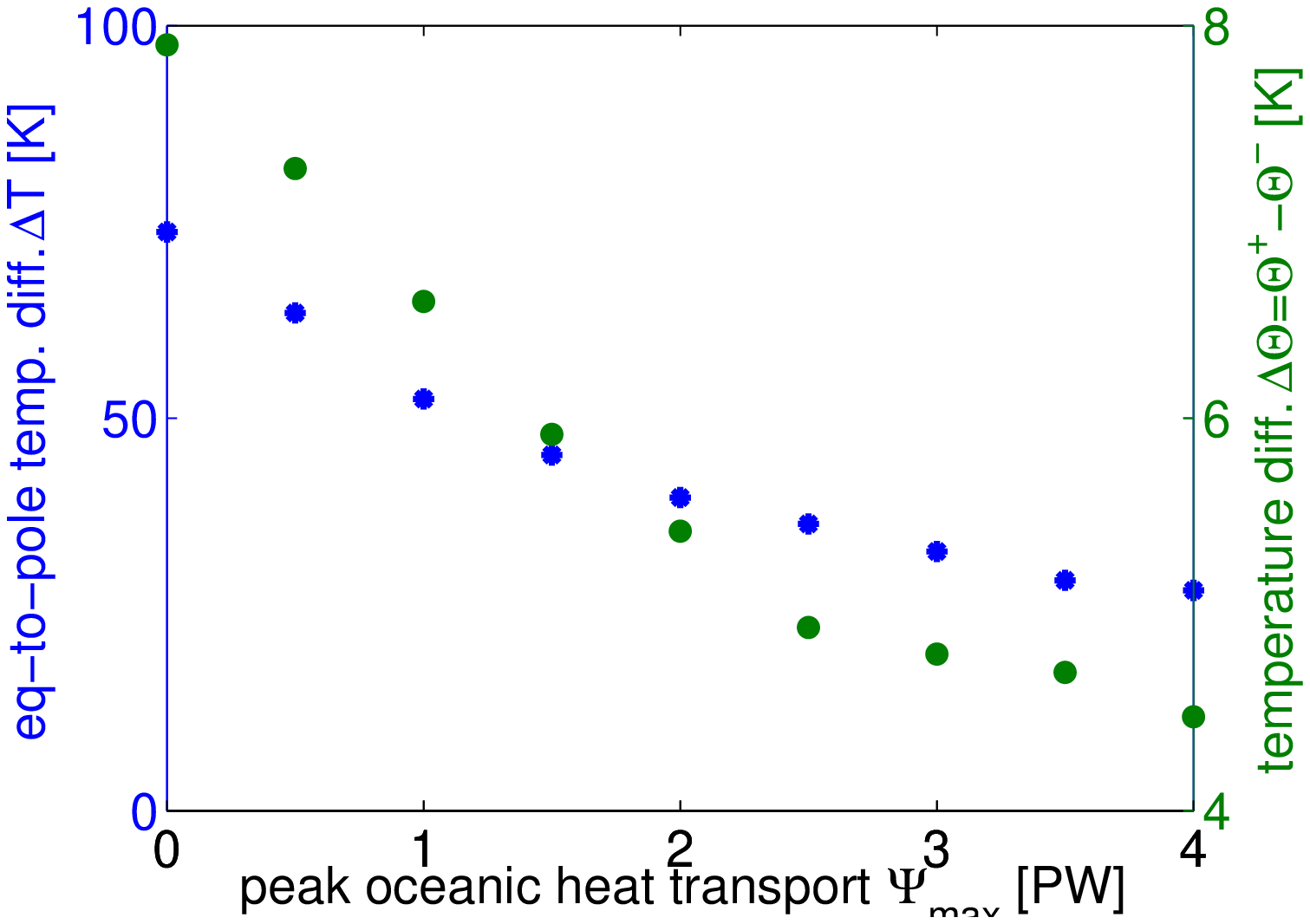}}\subfigure{\label{fig:reg2}\includegraphics[clip,height=150bp]{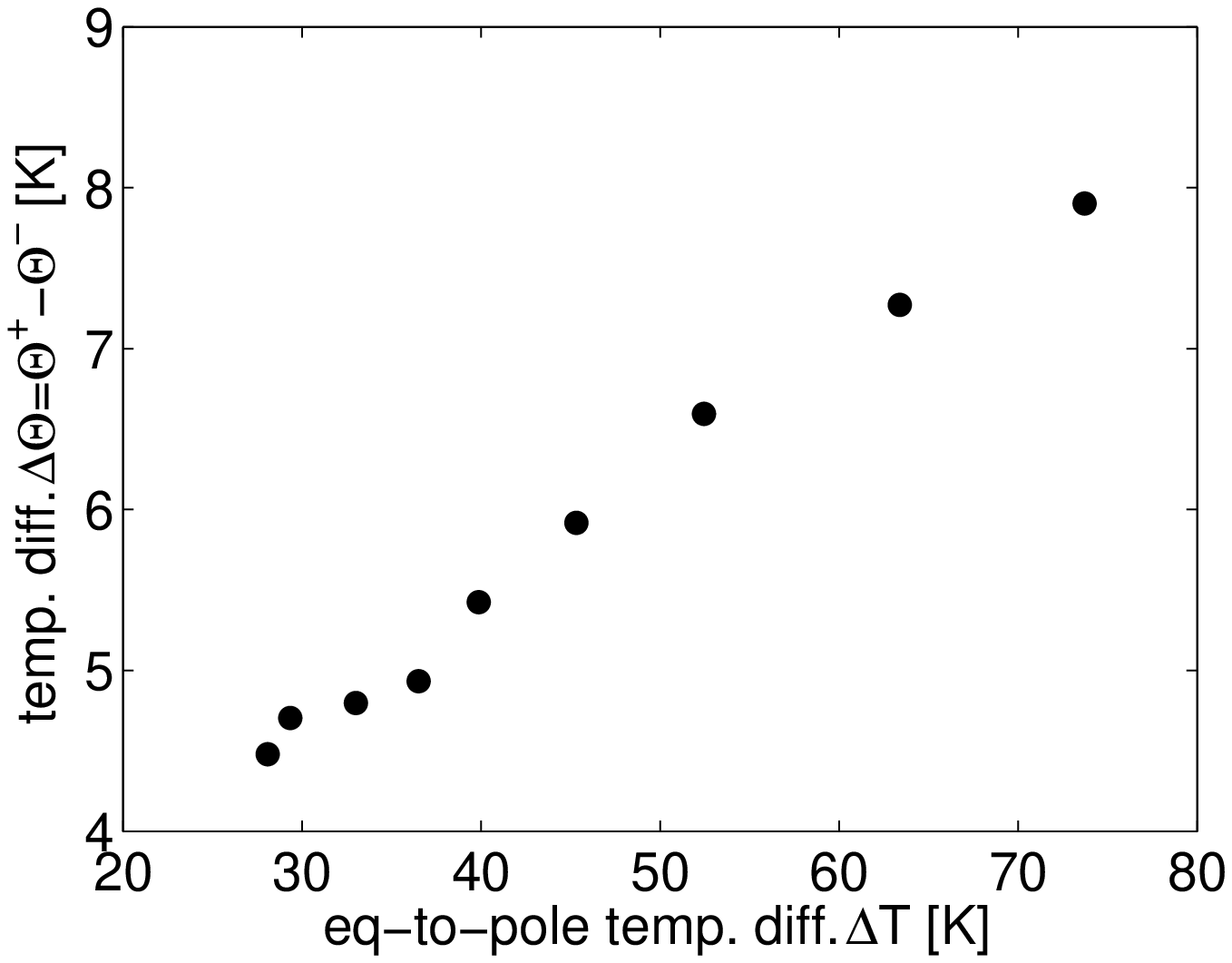}} 
\par\end{centering}

\caption{Scatter plot of time-averaged global mean temperature difference between
equator and pole (blue) as well as $\Theta^{+}$ and $\Theta^{-}$
(green) as a function of maximum energy transport in the ocean (a).
Scatter plot of time-averaged global mean temperature difference between
equator and pole as well as $\Theta^{+}$ and $\Theta^{-}$ (b). \label{fig:ta_temp_regress}}
\end{figure}

We observe on average a decline in $\Delta\Theta=\Theta^{+}-\Theta^{-}$
of approximately $0.4$\,K for every $0.5$\,PW increase in $\Psi_{\text{max}}$
(figure \ref{fig:reg1}; green graph). The total temperature difference
decreases from $7.9$\,K to $4.5$\,K across the considered range
of values of $\Psi_{\text{max}}$. The climate system becomes horizontally
more isothermal as $\Psi_{\text{max}}$ is reinforced, which is consistent
with the decline for the meridional difference in surface temperature
$\Delta T=T_{s,eq}-T_{s,pole}$ (figure \ref{fig:reg1}; blue graph).
We find an accurate linear relation between $\Delta T$ and $\Delta\Theta$
(the temperature difference between the two thermal reservoirs, $\Theta^{-}$
and $\Theta^{+}$): for every observed $10$\,K decline in $\Delta T$
the temperature difference, $\Delta\Theta$, decreases linearly by
approximately $0.8$\,K on average, as shown in figure \ref{fig:reg2}.
This provides a potentially interesting indication of how to relate
changes in the surface temperature gradient to quantities describing
the dynamic processes in the atmosphere.

\begin{onehalfspace}

\end{onehalfspace}

\begin{figure}[!t]
\begin{centering}
\subfigure{\label{fig:Effi}\includegraphics[clip,height=150bp]{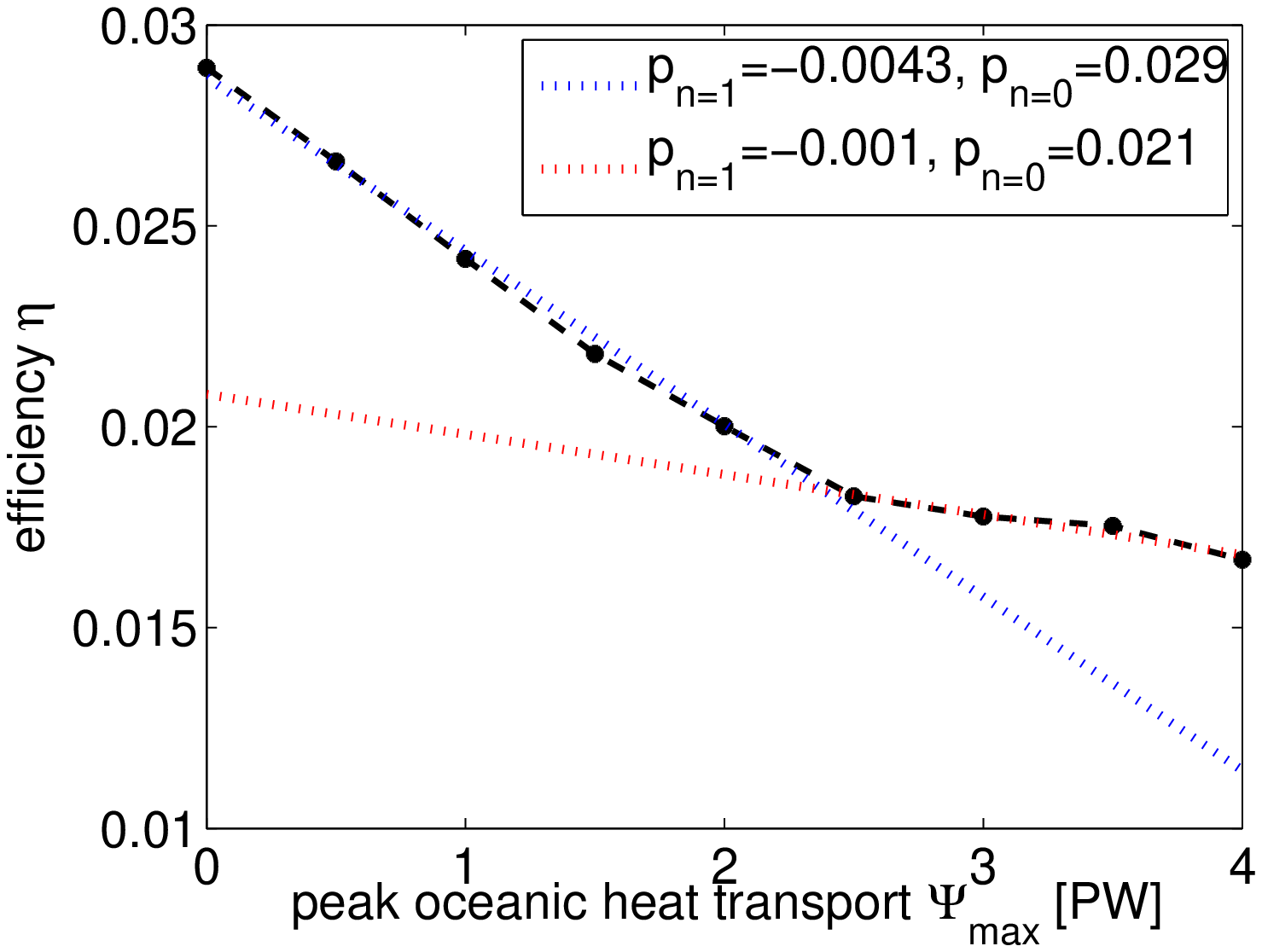}}\subfigure{\label{fig:W}\includegraphics[clip,height=150bp]{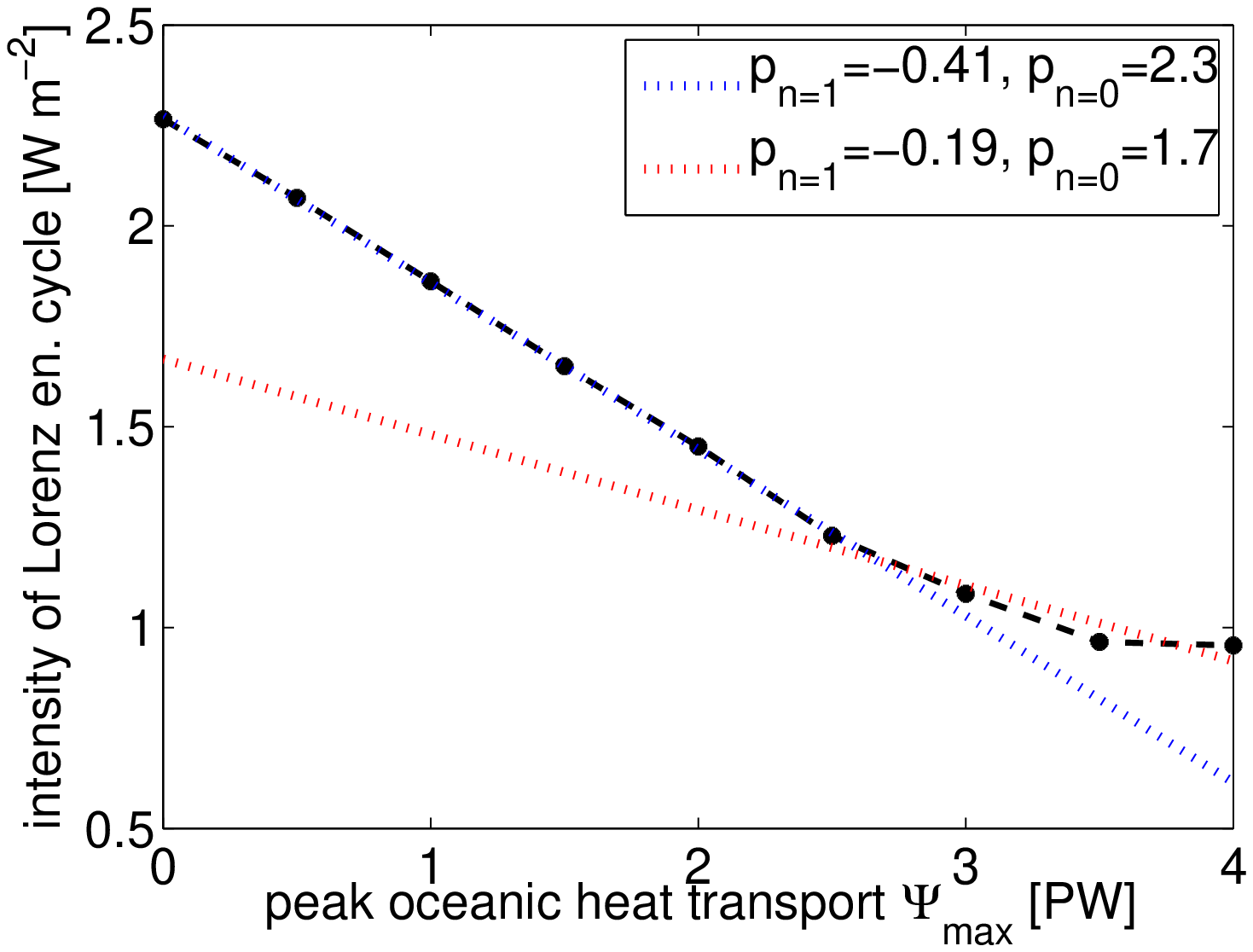}} 
\par\end{centering}

\caption{Time average of efficiency $\eta$ (a) and intensity of the Lorenz
energy cycle $\overline{\dot{W}}$ (b) for steady state obtained for
varying oceanic heat transport. Dotted line represents best linear
fit for i) $0.0\,\mathrm{PW}\leq\Psi_{\text{max}}\leq2.5$\,PW (blue)
and for ii) $2.5\,\mathrm{PW}\leq\Psi_{\text{max}}\leq4.0$\,PW (red)
with polynomial coefficients of n-th order, $\mathrm{p_{n=1}}$ and
$\mathrm{p_{n=0}}$. \label{fig:effi_work}}
\end{figure}

As the climate warms and the temperature difference between the warm
and the cold reservoir shrinks with increased $\Psi_{\text{max}}$,
the efficiency $\eta$ and the intensity of the Lorenz energy cycle
$\overline{\dot{W}}$ of the climate system decline (see figure \ref{fig:effi_work}).
The increase in $\Psi_{\text{max}}$ causes the climatic machine to
act less efficient, in terms of a decrease of the ratio between mechanical
energy output and thermal energy input.

\begin{onehalfspace}

Figure \ref{fig:effi_work} shows $\eta$ and $\overline{\dot{W}}$.
We observe a remarkably linear behaviour for both quantities when
considering the first 6 runs with $0.0\,\mathrm{PW}\leq\Psi_{\text{max}}\leq2.5$\,PW.
For every $0.5$\,PW increase in $\Psi_{\mathrm{max}}$ the efficiency
$\eta$ declines by about $2.0\cdot10^{-3}$, while the strength of
Lorenz energy cycle $\overline{\dot{W}}$ decreases by about $0.2$\,Wm$^{-2}$
(see dotted, blue graph in figure \ref{fig:effi_work}). For $\Psi_{\text{max}}$
larger than present-day values ($\Psi_{\text{max}}\leq2.5$), $\eta$
decreases by only $0.5\cdot10^{-3}$ per $0.5$\,PW increase, while
$\overline{\dot{W}}$ declines by $0.1$\,Wm$^{-2}$ per $0.5$\,PW
increase (see dotted, red graph in figure \ref{fig:effi_work}). We
observe an abrupt change in the tendency for $\Psi_{\text{max}}=2.5$;
at which pronounced tropical and subtropical atmospheric cooling sets
in. This indicates that the change in the temperature difference between
equatorial and tropical regions cause a drastic change in the dynamical
properties of the system.

The reason for this enhanced decrease in $\overline{\dot{W}}$ can
be found in the decrease of the temperature difference between the
warm and the cold reservoir. From energy conservation we know, the
decrease in the strength of Lorenz energy cycle $\overline{\dot{W}}$
implies that also the total dissipation $\overline{\dot{D}}$ decreases
in a steady state climate, as the climatic engine has smaller rate
of transformation of available into kinetic energy. The decrease of
$\overline{\dot{D}}$ implies, e.g. that surface winds are weaker,
because this is where most of the dissipation takes place. We note
that by increasing $\Psi_{\text{max}}$, warm and cold air masses
get mixed more effectively with the result that the atmosphere becomes
horizontally more isothermal and, hence, the climatic engine acts
less efficiently.

\begin{figure}[!t]
\subfigure{\label{fig:Entropy}\includegraphics[clip,height=150bp]{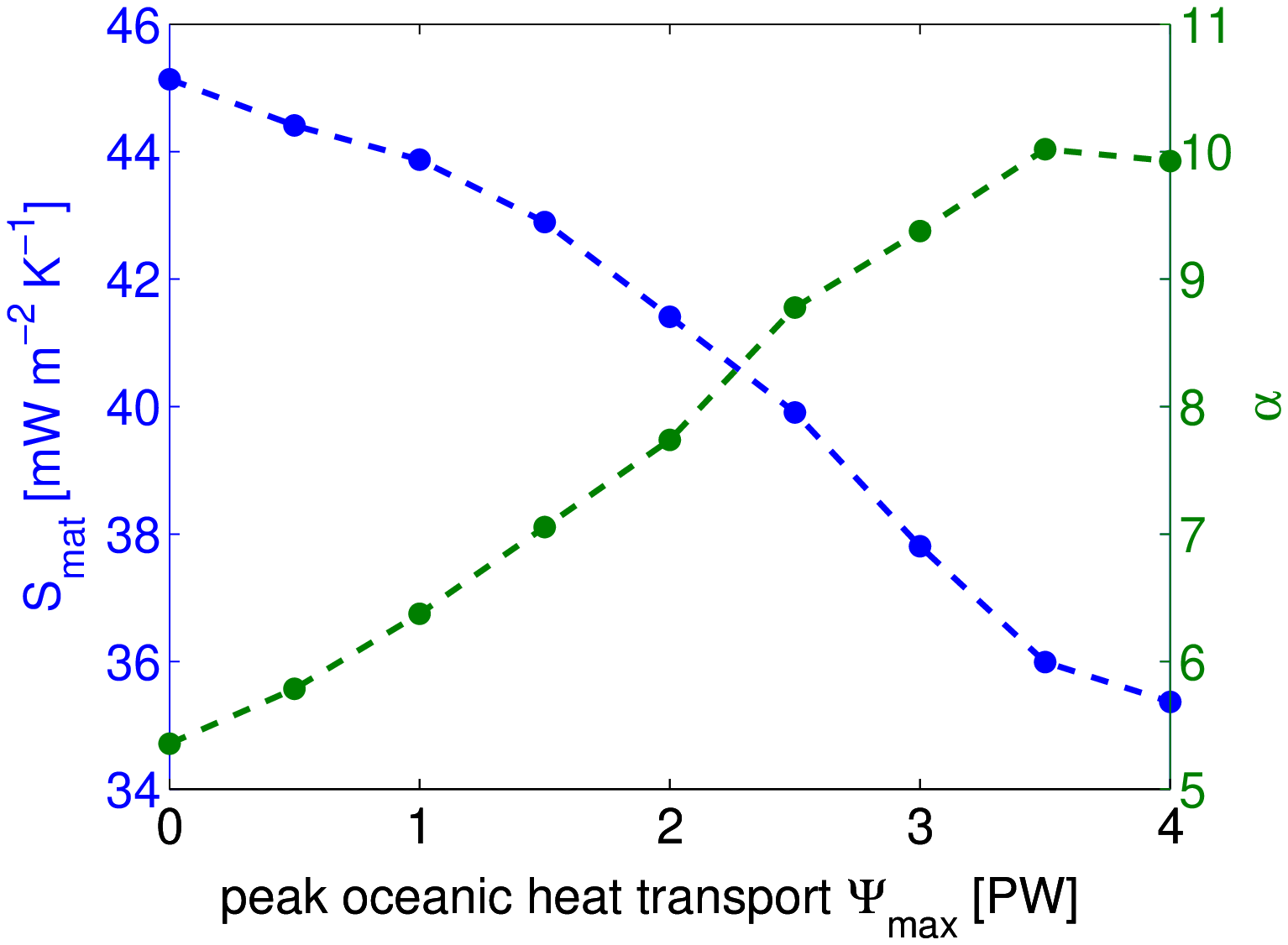}}\subfigure{\label{fig:En-dist}\includegraphics[clip,height=150bp]{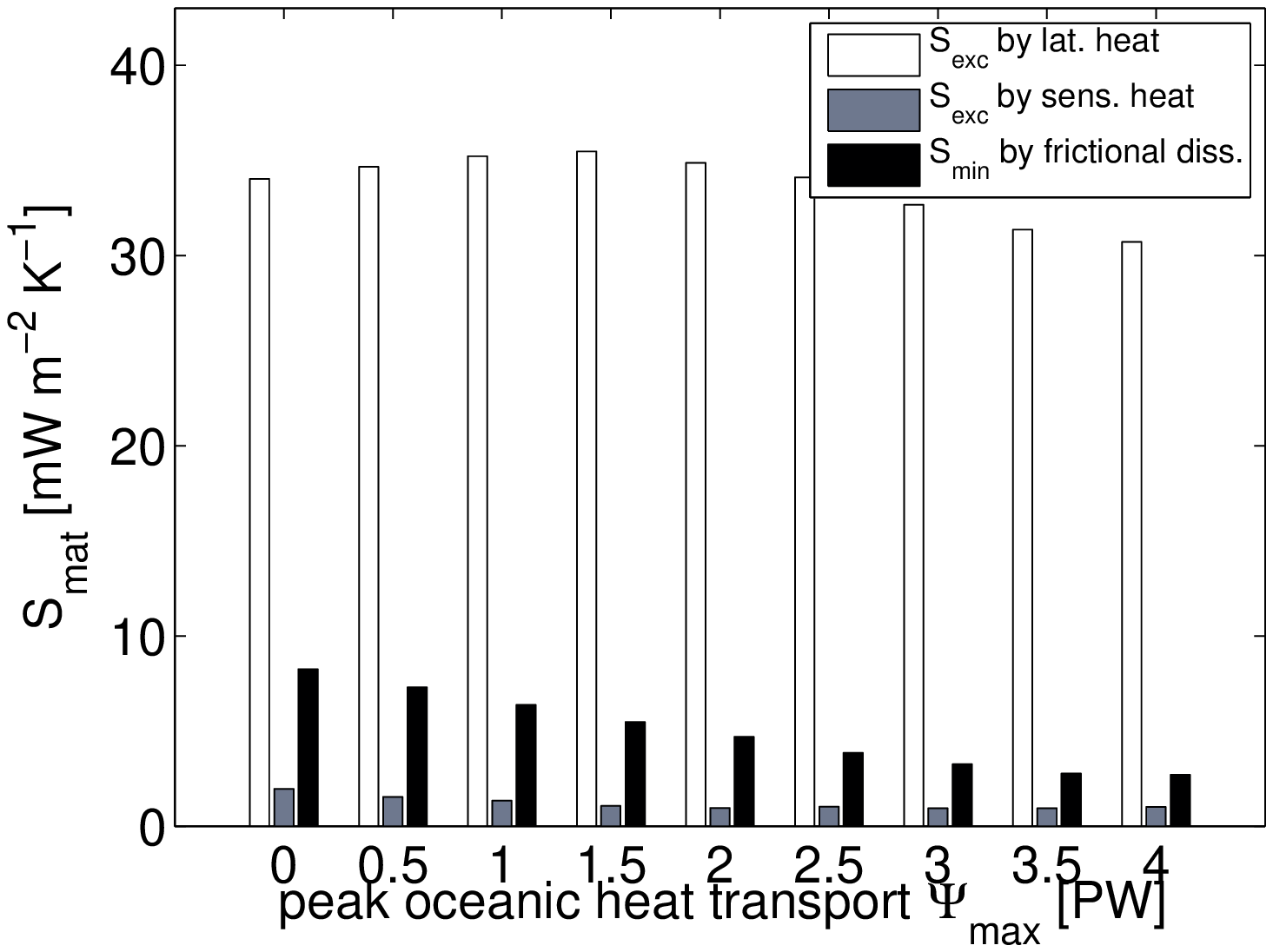}}

\caption{In (a) steady-state global mean material entropy production $\overline{\dot{S}_{mat}}$
(blue graph) and degree of irreversibility $\alpha$, and in (b) most
relevant contributions of $\overline{\dot{S}_{mat}}$ split into $\overline{\dot{S}_{exc}}$
and $\overline{\dot{S}_{min}}$, as a function of increasing oceanic
heat transport. \label{fig:entropy_contributions}}
\end{figure}

Material entropy production $\overline{\dot{S}_{mat}}$, introduced
in equation \ref{eq:mat-entropy}, as well as the degree of irreversibility
$\alpha$, introduced in equation \ref{eq:alpha}, are shown in figure
\ref{fig:Entropy}. With increasing values of $\Psi_{\text{max}}$,
the decrease in the intensity of the Lorenz energy cycle and the increase
in the surface temperature imply a reduction of the part in $\overline{\dot{S}_{mat}}$
linked with frictional dissipation, which is related to lower bound
of entropy production $\overline{\dot{S}_{min}}$. Nonetheless, one
needs to investigate the excess of entropy production $\overline{\dot{S}_{exc}}$,
which is linked to the turbulent heat fluxes down the temperature
gradient. The relative decrease in entropy production due to frictional
dissipation ($\overline{\dot{S}_{min}}$) is stronger than the relative
decrease in entropy production by down-gradient turbulent heat transport
($\overline{\dot{S}_{exc}}$) as featured by the overall increase
in $\alpha$ (figure \ref{fig:Entropy}). Thus, the entropy production
due to the turbulent heat transport down the gradient of the temperature
field becomes more and more dominant as the oceanic transport increases
because irreversible mixing becomes stronger. As a result the degree
of irreversibility increases since larger oceanic heat transport implies
larger mixing which impels $\overline{\dot{S}_{exc}}$.

In figure \ref{fig:En-dist} the main contributions of the material
entropy production in the model are displayed. This includes the contributive
processes due to latent and sensible turbulent heat fluxes and frictional
dissipation of kinetic energy. Entropy production by latent heat,
including convective as well as large-scale precipitation, surface
latent heat fluxes and rain-snow conversion processes, makes by far
the largest portion of material entropy production. For small intensities
of $\Psi_{\text{max}}$, the value of entropy production by latent
heat reads $35$\,mW\,m$^{-2}$\,K$^{-1}$. For increasing $\Psi_{\text{max}}$
up to $1.5$\,PW the value increases by $2$\,mW\,m$^{-2}$\,K$^{-1}$,
while for larger values of $\Psi_{\text{max}}$, this contribution
to entropy production declines by $4$\,mW\,m$^{-2}$\,K$^{-1}$.
Entropy production by frictional dissipation decreases, as one would
expect since dissipation is proportional to the intensity of the Lorenz
energy cycle (shown in figure \ref{fig:W}). It decreases from $8$\,mW\,m$^{-2}$\,K$^{-1}$
for $\Psi_{\text{max}}=0$\,PW down to $3$\,mW\,m$^{-2}$\,K$^{-1}$
for $\Psi_{\text{max}}=4$\,PW. Entropy production by sensible turbulent
heat flux at the surface as well as in the atmosphere decreases by
half (from $2$\,mW\,m$^{-2}$\,K$^{-1}$ to $1$\,mW\,m$^{-2}$\,K$^{-1}$)
with $\Psi_{\text{max}}$ increasing. One would expect that larger
values of $\Psi_{\text{max}}$ would lead to larger values of $\overline{\dot{S}_{mat}}$,
using the argument that a warmer planet should be able to have a stronger
hydrological cycle. For low values of $\Psi_{\text{max}}$, the increase
in $\overline{\dot{S}_{mat}}$ due to the hydrological cycle is overcompensated
by the decrease in the contribution due to the frictional dissipation.

\begin{figure}[H]
\begin{centering}
\subfigure{\label{fig:EP-1}\includegraphics[height=165bp]{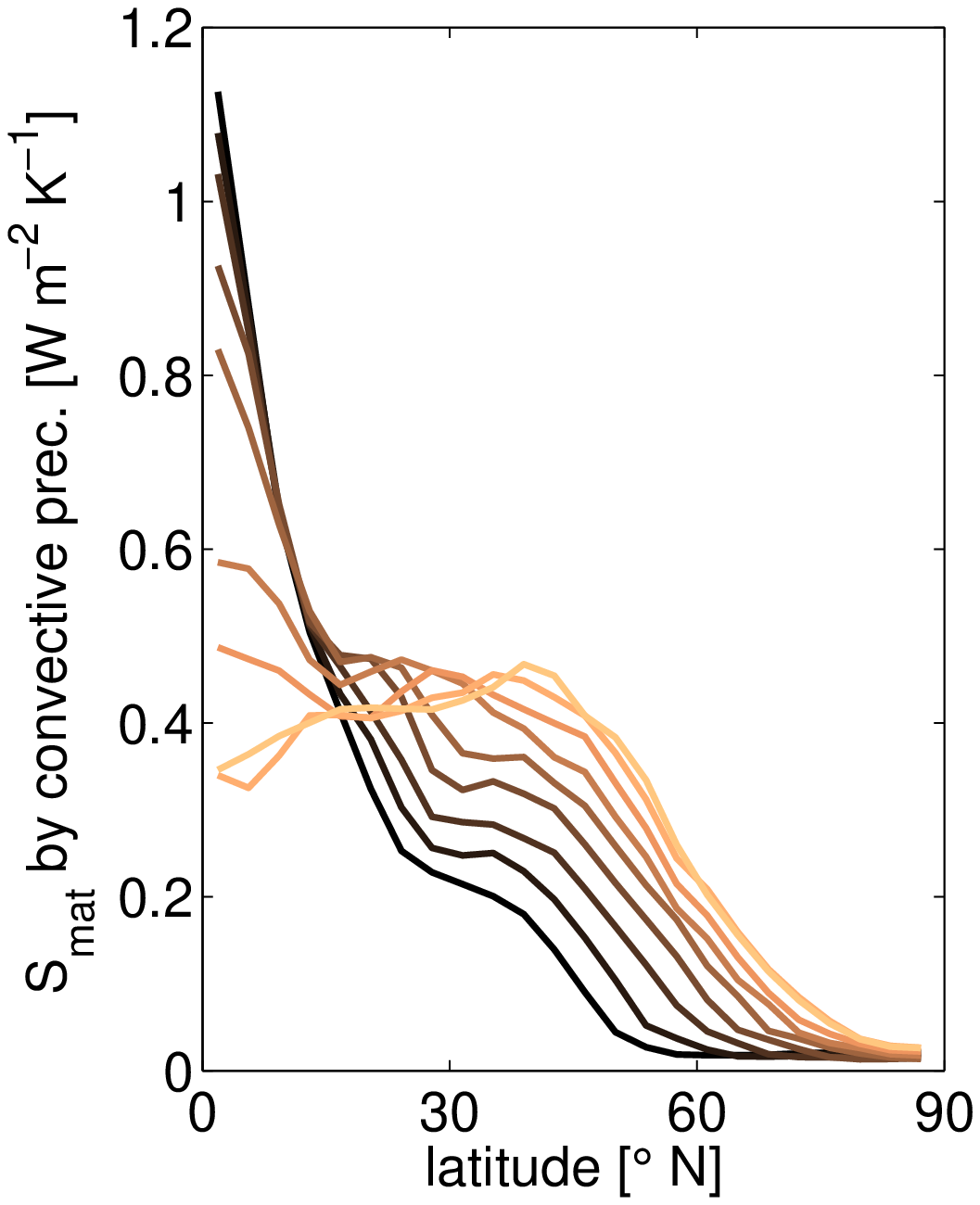}}~~~~\subfigure{\label{fig:EP-2}\includegraphics[height=165bp]{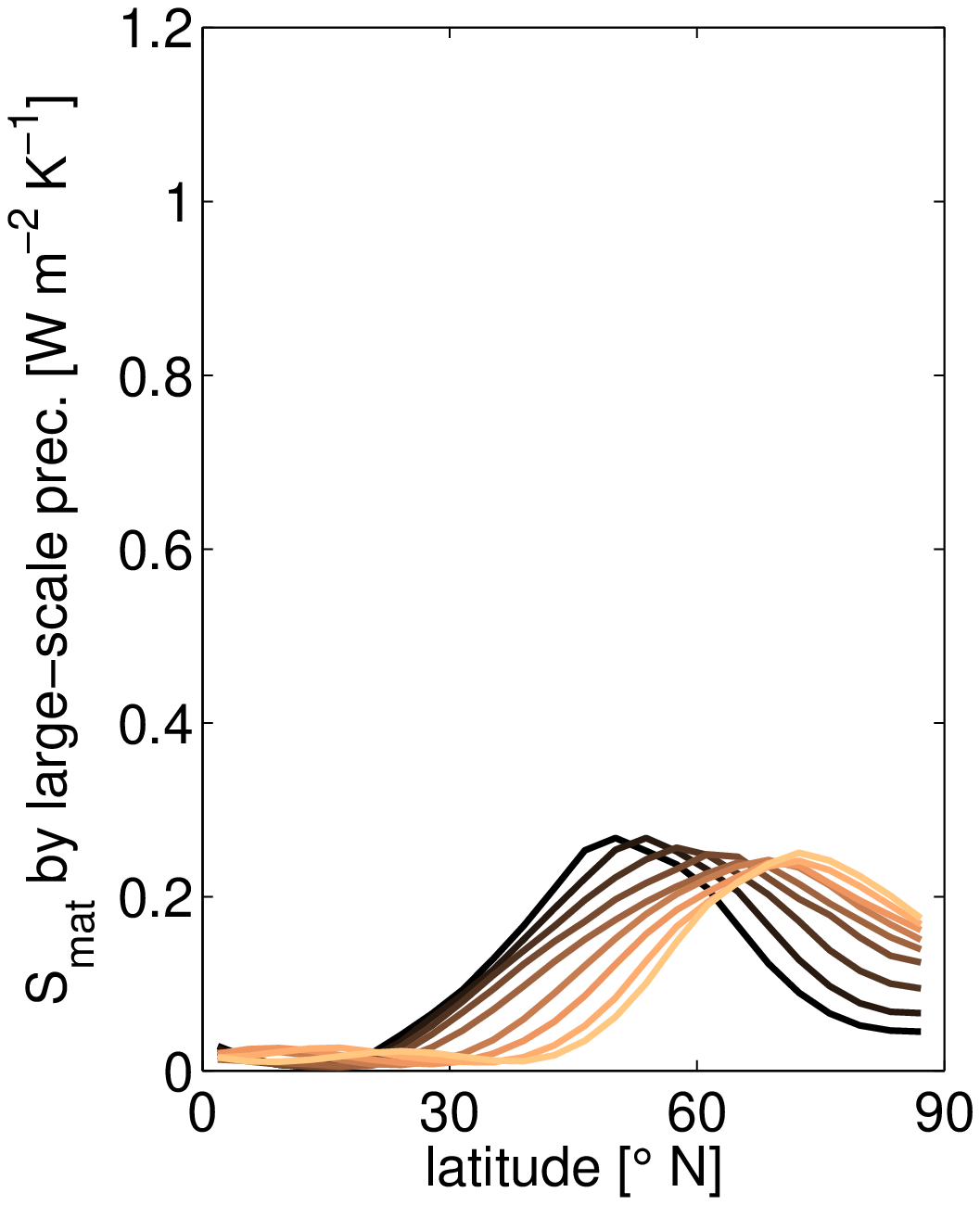}}~~~~\subfigure{\label{fig:EP-3}\includegraphics[height=165bp]{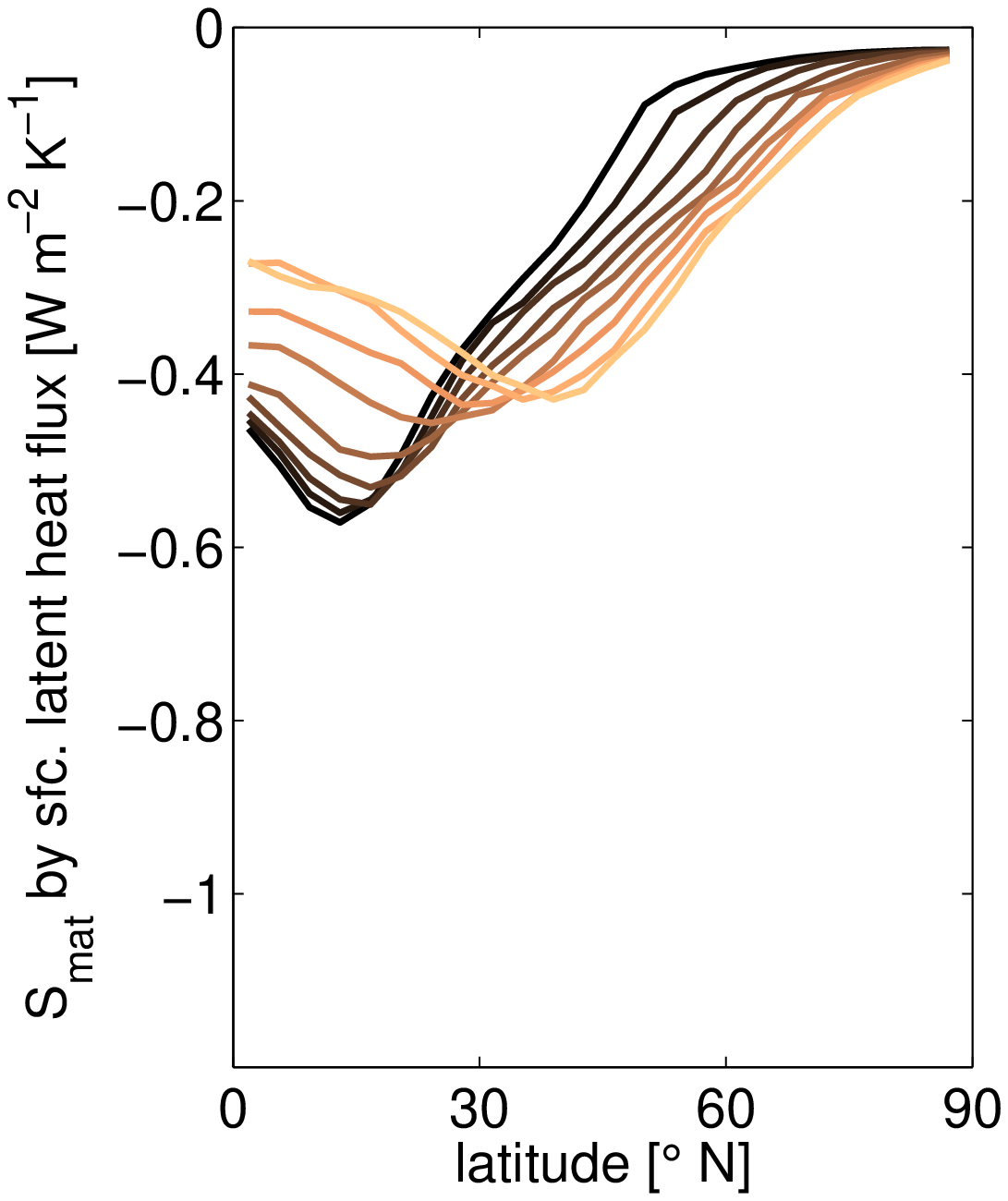}}\linebreak{}
 \subfigure{\label{fig:EP-4}\includegraphics[height=165bp]{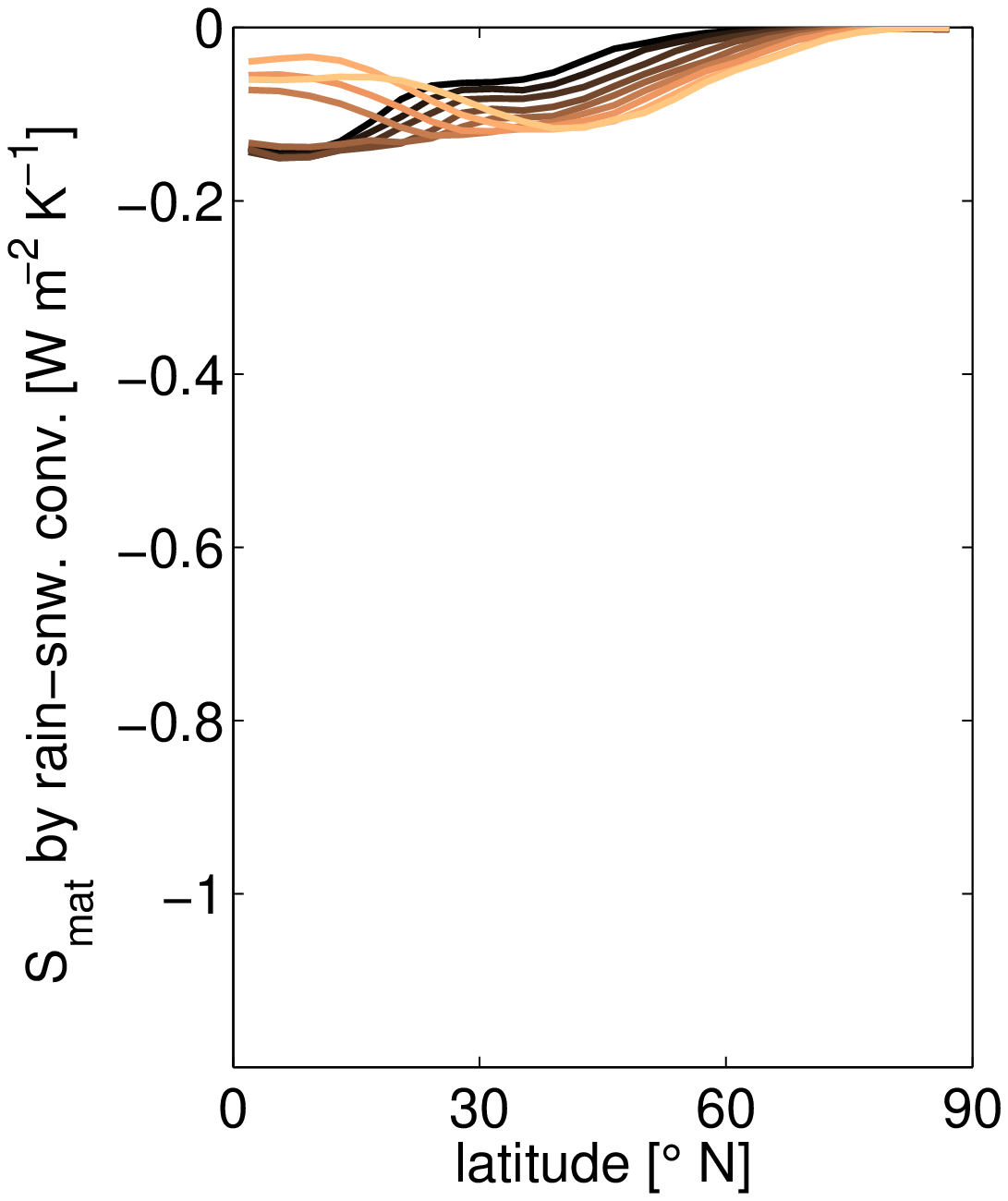}}~~~~\subfigure{\label{fig:EP-5}\includegraphics[height=165bp]{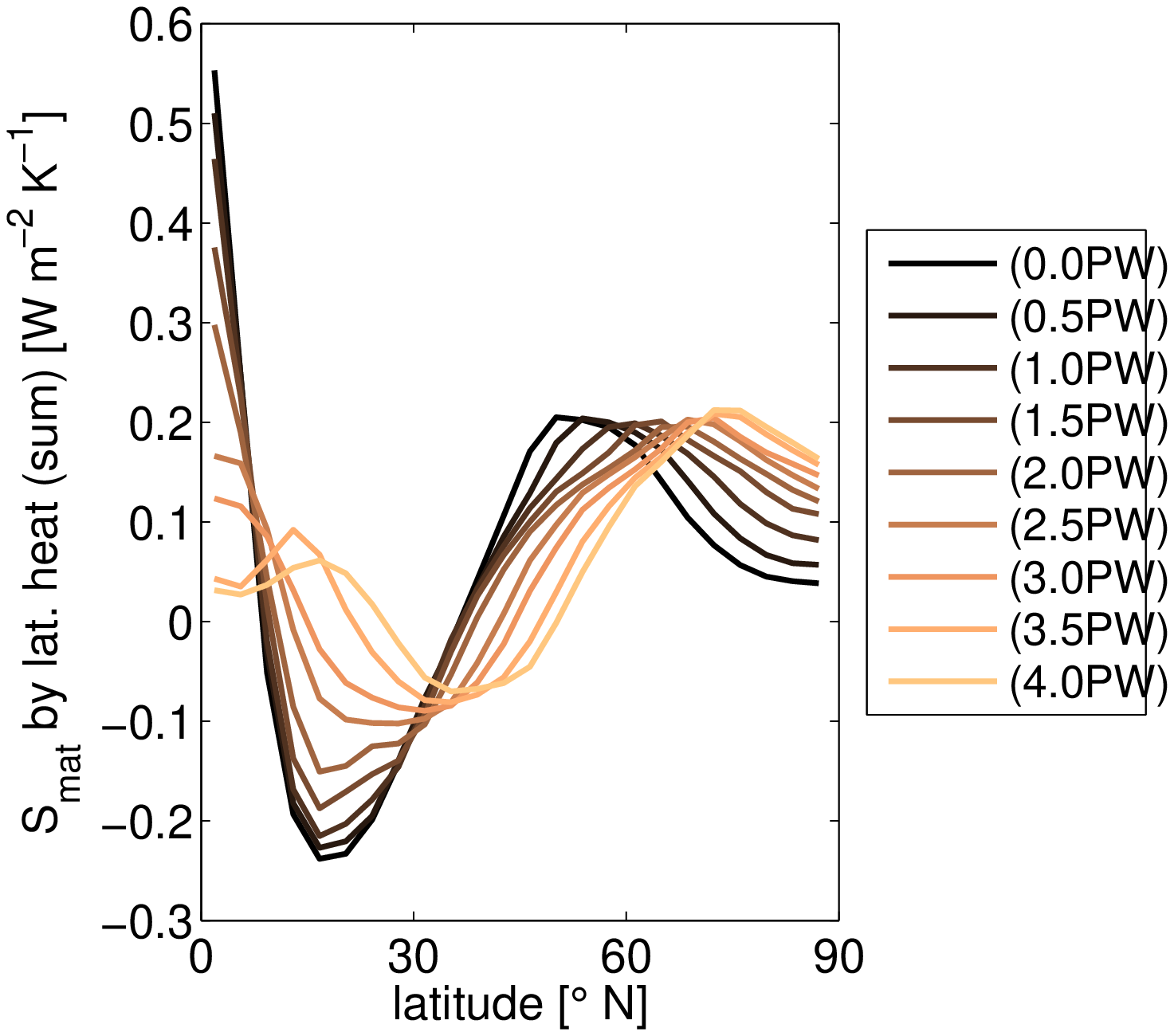}} 
\par\end{centering}

\caption{Time averaged zonal mean of 4 contributions of material entropy production
concerning latent heat processes as a function of increasing oceanic
heat transport. Material entropy production by convective precipitation
(a); by large-scale precipitation (b); by surface latent heat fluxes
(c); by rain-snow conversion (d). Sum of these 4 contributions (e).
\label{fig:entropy_latent_contributions}}
\end{figure}

In order to further clarify the impacts on the material entropy production
of increasing $\Psi_{\text{max}}$ , we split the material entropy
production due to irreversible latent turbulent heat processes into
the contributions coming from convective precipitation, large-scale
precipitation, surface latent heat fluxes, and heat release by rain-snow
conversion. Figure \ref{fig:EP-1}-\ref{fig:EP-4} displays the time
and zonal mean of these 4 contributions, where the process implicating
convective precipitation gives the largest contribution, particularly
in the tropics and subtropics, where most of atmospheric convection
processes takes place. Note that the divergence of the horizontal
turbulent latent and sensible heat fluxes divided by the local temperature,
which correspond to the boundary part in equation \ref{eq:mat-entropy}
(the other being the surface fluxes of latent heat) are negligible. 
\end{onehalfspace}

We observe that the peak at the equator is significantly reduced,
while convection processes move into the mid-latitudes for increased
$\Psi_{\text{max}}$ where the surfaced is heated and static stability
decreases. Large-scale precipitation features are shifted out of the
mid-latitudes towards higher latitudes. As large-scale precipitation
regimes experience a shift to higher latitudes, their maximum intensity
is almost kept constant. For the control run, latent heat fluxes at
the surface show a maximum at latitudes of $20\text{\textdegree}$
to $25\text{\textdegree}$. These maxima on both hemispheres indicate
the region with maximum evaporation. As the heat transport in the
ocean is increased, latent turbulent heat fluxes reduces largely in
tropical and subtropical regions, and peak latent heat fluxes move
towards mid-latitudes. The region with largest evaporation at the
surface shifts from the subtropics to the mid-latitudes with increasing
$\Psi_{\text{max}}$. Atmospheric latent heat release by rain-snow
conversion qualitatively shows, as expected due to high atmospheric
processes, similar patterns as the meridional profile for convective
processes. Tropical regions are characterised by a considerable reduction
of heat release, whereas the subtropics and mid-latitudes gain heat
due to rain-snow conversion.

The material entropy production is negative in the tropical latitudinal
band (figure \ref{fig:EP-5}). This, of course, is perfectly compatible
with the second law of thermodynamics and results from the fact that
there is a net large scale transport of energy from those regions
to the equator and to the mid latitudes as result of net moisture
transport (figure \ref{fig:AHT}). Such a negative contribution is
overcompensated by the positive material entropy production associate
to the absorption of the transported latent heat.

\begin{onehalfspace}

\end{onehalfspace}

\section{Conclusions}

In this investigation we have studied the climate sensitivity to changes
in the ocean heat transport describing the response of macroscale
thermodynamical properties of the climate system, coming from a theoretical
framework introduced by \citet{Lucarini2009}. \citet{Stone1977}
states a full oceanic-atmospheric compensation, so that the total
heat transport, as the sum of the oceanic and the atmospheric component,
is insensitive to internal parameters of the atmosphere-ocean system,
e.g. the meridional temperature gradient.

Our results show an almost exact compensation of the meridional heat
transport in the atmosphere for increasing the meridional oceanic
heat transport,$\Psi_{\text{max}}$, from $0.0$\,PW to $4.0$\,PW
by $0.5$\,PW. As a result, the total heat transport is almost insensitive
to the direct impact of increasing the oceanic heat flux following
\citet{Stone1977}. Minor deviations are assumed to stem from an altered
planetary albedo due to a retreat in sea ice at the poles and a relocation
of cloud formations.

\begin{onehalfspace}
Two major features in temperature properties of the climate system
on the increase of meridional heat transport ($\Psi_{\text{max}}=0.0$\,PW
$\rightarrow$ $\Psi_{\text{max}}=4.0$\,PW) in the ocean can be
noted: i) an increase in global mean surface temperature $T_{s}$
of $10$\,K; ii) the reduction in the difference between the warm
pool temperature $\Theta^{+}$ and the cold pool temperature $\Theta^{-}$.
The latter implies that the atmospheric system becomes more isothermal
in the horizontal with increasing oceanic heat transport. Main cause
for this reduction of the global temperature gradient is the enhancement
of convergence of latent heat fluxes. Convection spreads out from
the deep tropics into the mid-latitudes leading to a large range of
dynamic and thermodynamic changes. Warming of the troposphere in the
mid- and high latitudes results in increased water vapour content.
The results indicate that the investigated system becomes less efficient
and more irreversible while planetary entropy production declines
for increasing strength of the oceanic heat transport. The effect
of thermalisation leading to the reduction of the efficiency of the
system with increasing intensity of the ocean heat transport can be
related to the decrease in the reservoir of the potential energy available
for conversion in the Lorenz energy cycle (not show). The intensity
of the Lorenz energy cycle declines by $1.3$\,Wm$^{-2}$, while
material entropy production reduces by $9.8$\,mWm$^{-2}$K$^{-1}$
(both values for increasing $\Psi_{\text{max}}=0.0$\,PW $\rightarrow$
$\Psi_{\text{max}}=4.0$\,PW).

The temperature difference between the warm ($\Theta^{+}$) and the
cold ($\Theta^{-}$) heat reservoir decreases by $3.4$\,K in total
for increasing oceanic heat transport. This is basically caused by
an enhanced warming of the polar and subpolar low troposphere as well
as tropospheric cooling above the tropics. This warming is mainly
due to intensified latent heat release by large-scale precipitation
as corresponding heating patterns experience a shift from the mid-
to high latitudes in the course of increasing the oceanic heat transport.
This will lead to further warming due to the water vapour feedback
\citep{Herweijer2005,Barreiro2011a}.

By the result of reducing the planetary thermal difference the amount
of available potential energy is decreased, leading to a decline in
the transformation of kinetic energy. The intensity of the Lorenz
energy cycle, thus, is reduced with the effect that the Hadley and
the Ferrel circulation regimes experience a latitudinal shift and
a decrease in intensity when increasing the heat transport in the
ocean (not shown). 
\end{onehalfspace}

With increasing oceanic heat transport the surface temperature in
the mid-latitudes rises. This surface heating destabilises the low-tropospheric
air masses which respond with enhanced convective processes in the
mid-latitude. As more heat is taken up by the ocean in the tropics
and more heat is released in the mid-latitudes, heat is taken out
of the Hadley cell regime, being most active in the tropics and subtropics,
and is then released in the storm track area in the mid-latitudes
\citep{Rose2013}.

\begin{onehalfspace}
As the atmosphere becomes more isothermal due to intensified mixing
when increasing the oceanic heat transport, the strength of the Lorenz
energy cycle as well as the efficiency decreases. This is consistent
with the results in \citet{Lucarini2010} where higher $\mathrm{CO}{}_{2}$
concentrations in the atmosphere result in global warming and smaller
temperature differences in the atmosphere with a resulting decline
in the efficiency as well as in the intensity of the Lorenz energy
cycle. 
\end{onehalfspace}

When considering stronger oceanic transport, the climate system is
characterised by a declining total material entropy production, while
the degree of irreversibility increases, since the decrease in entropy
production by frictional dissipation is more intense than the decrease
in entropy generation due to sensible and, in particular, latent heat
flux. The flux of latent heat contributes most to the material entropy
production in the climate system. When increasing the heat transport
in the ocean from $0.0$\,PW to $1.5$\,PW, material entropy production
due to latent heat flux increases which can be explained by an outspread
of convection from the deep tropics into the mid latitudes, while
the maximum latent release is still located in the central tropics.
When increasing the heat transport further, convective processes collapse
in the deeps tropics and, thus, affecting evaporation intensities
at tropical sea surface by reducing it. As a result, a decrease in
material entropy production by latent heat fluxes can be noted from
the increase of the oceanic heat transport larger than $2.0$\,PW.

\begin{onehalfspace}
In order to broaden the outlook in the field of study concerning the
role of oceanic heat transport, one could investigate the role of
latitudinal location of the peak oceanic heat transport on macroscale
thermodynamic properties. In the present set of experiments the peak
oceanic transport was fixed at the latitude of 27\textdegree{}. The
location of the peak can be adjusted by altering $N$ in equation
\ref{eq:oht-rose}. This would complement the investigation by \citet{Rose2013}
and help understanding the properties of warm equable climates. One
could apply the same theoretical framework as being used in this study.
Another possible future line of investigation deals with studying
planets with different astrophysical parameters, such as rotation
rate, eccentricity, and obliquity, with the goal of contributing to
the rapidly growing field of investigation of the atmosphere of exoplanets. 
\end{onehalfspace}

\addsec{Acknowledgement}

The authors acknowledge the support by the DFG Cluster of Excellence
CLISAP. VL acknowledges the support of the FP7-ERC Starting Investigator
Grant NAMASTE - Thermodynamics of the Climate System (No. 257037).

\begin{onehalfspace}
\bibliographystyle{apalike} \phantomsection\addcontentsline{toc}{section}{\refname}\bibliographystyle{apalike}
\bibliography{20141009_ClimateDynamics}
 \end{onehalfspace}

\end{document}